\documentclass[12pt,authoryear]{elsarticle}

\usepackage[top=2cm,bottom=2.65cm,left=2.65cm,right=2.55cm]{geometry}
\usepackage{pslatex}

\usepackage{textpos}
\usepackage{caption}
\usepackage{tcolorbox}
\usepackage{algorithm}
\usepackage{algpseudocode}
\usepackage{listings}
\usepackage{color,xcolor}
\usepackage{textcomp}
\usepackage{pslatex}
\usepackage{graphicx}

\usepackage{epsfig}
\usepackage{epstopdf}
\usepackage{pgf}

\usepackage{amssymb,amsthm,amsfonts,amsmath,bm,amsbsy,mathtools,bbm,relsize}

\usepackage{enumitem}
\usepackage{mathtools}
\usepackage{longtable}

\usepackage{verbatim}
\usepackage{float}
\usepackage{appendix}
\usepackage{array}
\usepackage{varwidth}
\usepackage{multirow}
\usepackage{tabularx}
\usepackage{framed}
\usepackage{lipsum}
\usepackage[round]{natbib}
\usepackage[hidelinks]{hyperref}

\usepackage{indentfirst}
\usepackage{longtable}
\usepackage[utf8]{inputenc}
\usepackage{booktabs}
\usepackage{lscape}
\usepackage{rotating}
\usepackage{fancyvrb}
\usepackage[caption = false]{subfig}
\usepackage{setspace}
\usepackage[T1]{fontenc}

\usepackage{lineno}
\usepackage{mathtools}
\usepackage{bigints}
\usepackage{url}
\usepackage{bm,bbm}
\usepackage[normalem]{ulem}

\usepackage{pzccal}
\DeclareFontFamily{OT1}{pzc}{}
\DeclareFontShape{OT1}{pzc}{m}{it}{<-> s * [1.10] pzcmi7t}{}
\DeclareMathAlphabet{\mathcal}{OT1}{pzc}{m}{it}

\newtheorem{theorem}{Theorem}
\newtheorem{lemma}[theorem]{Lemma}	
\newtheorem{proposition}[theorem]{Proposition}	
\newtheorem{corollary}[theorem]{Corollary}	
\newtheorem{remark}{Remark}

\newcommand*{\LongState}[1]{\State
\parbox[t]{\linewidth-\algorithmicindent}{#1\strut}}
	
\newcolumntype{"}{@{\vrule width 1pt}}

\def\infinity{\rotatebox{90}{8}}

\newcommand{\pkg}[1]{\texttt{#1}}

\def\rset{\mathbb{R}}
\def\param{\theta}
\def\paramspa{\Theta}
\def\targ{\pi}
\def\esp{\mathbb{E}}
\def\var{\mathbb{V}}
\def\models{\mathcal{M}}

\def\htheta{\hat{\theta}}

\def\rset{\mathbb{R}}

\allowdisplaybreaks

\journal{}
\begin{document}
\begin{frontmatter}

\title{Model comparison for Gibbs random fields using \\noisy reversible jump Markov chain Monte Carlo}

\author{Lampros Bouranis}
\cortext[corresp]{Corresponding author}
\ead{lampros.bouranis@ucdconnect.ie}

\author{Nial Friel\corref{corresp}}
\ead{nial.friel@ucd.ie}

\author{Florian Maire}
\ead{florian.maire@ucd.ie}

\address{School of Mathematics and Statistics \& Insight Centre for Data Analytics, \\ University College Dublin, Ireland}

\begin{abstract}
\noindent
The reversible jump Markov chain Monte Carlo (RJMCMC) method offers an across-model simulation approach for Bayesian estimation and model comparison, by exploring the sampling space that consists of several models of possibly varying dimensions. 
A naive implementation of RJMCMC to models like Gibbs random fields suffers from computational difficulties: the posterior distribution for each model is termed doubly-intractable since computation of the likelihood function is rarely available. Consequently, it is simply impossible to simulate a transition of the Markov chain in the presence of likelihood intractability. 
A variant of RJMCMC is presented, called noisy RJMCMC, where the underlying transition kernel is replaced with an approximation based on unbiased estimators. 
Based on previous theoretical developments, convergence guarantees for the noisy RJMCMC algorithm are provided.
The experiments show that the noisy RJMCMC algorithm can be much more efficient than other exact methods, provided that an estimator with controlled Monte Carlo variance is used, a fact which is in agreement with the theoretical analysis.
\end{abstract}

\end{frontmatter}

\section{Introduction}\label{section:intro}
Model selection is a problem of great importance in statistical science. The aim is to choose which model among a set of possible ones best describes the data $y\in \mathcal{Y}$. 
From a Bayesian perspective, the prior beliefs for each model are reflected by a prior distribution and this information is then updated subjectively when data are observed. This step is typically carried out by calculating the marginal likelihood or \textit{evidence} for each model, which is defined as the integrated likelihood with respect to the prior measure. In many cases, this quantity cannot be derived analytically and thus needs to be estimated.

This paper considers the problem of Bayesian model comparison of \textit{doubly-intractable} distributions. 
The motivating application is Gibbs random fields (GRFs), which are discrete-valued Markov random fields where an intractable normalising constant that depends on the model parameters, $z(\theta)$, is present for the tractable un-normalised likelihood $q(y\mid\theta)$. 
The likelihood density, given a vector of parameters $\param\in\paramspa\subseteq\mathbb{R}^{d}$ and a vector of statistics $s(y)\in \mathcal{S}\subseteq\mathbb{R}^{d}_{+}$ that are sufficient for the likelihood, is
\begin{align} \label{intro_eqn:likelihood}
f(y\mid\theta)=\frac{q(y\mid\theta)}{z(\theta)}=\frac{\exp\left\{\param^\top s(y)\right\}} {\sum_{y\in \mathcal{Y}}^{} \exp\left\{\param^\top s(y)\right\}}.
\end{align}
Posterior parameter estimation for GRFs, which have found applicability in areas like image analysis and disease mapping \citep{rue}, genetic analysis \citep{francois} and social network analysis \citep{wasserman}, is termed a doubly-intractable problem because the normalisation terms of both the likelihood function and the posterior distribution $\pi(\param\mid y)\propto f(y\mid\theta)p(\param)$ are intractable. 
Bayesian model comparison of such models has attracted the attention of researchers and several methods have been considered, relying on likelihood simulations \citep{friel4,caimo3, johansen}, approximations to the intractable likelihood \citep{bouranis2} and likelihood-free simulation techniques like the Approximate Bayesian Computation (ABC) algorithm \citep{grelaud2009}
.

In this paper we explore trans-dimensional Markov chain Monte Carlo (MCMC) for GRFs, focusing on the direct approach of a single across-model Markov chain using the celebrated reversible jump MCMC (RJMCMC) technique \citep{green95}. 
The advantage of across-model approaches is that they avoid the need for computing the evidence for each competing model by treating the model indicator as a parameter, where the chain explores simultaneously the model set and the parameter space. 
In the context of GRFs, however, RJMCMC techniques simply cannot be implemented because the likelihood normalising constant $z(\theta)$ cannot be computed point-wise. Below we present a summary of the main contributions of this paper.

A variant of the RJMCMC algorithm is developed, where the intractable ratio of normalising constants that the acceptance probability depends on is approximated by an unbiased estimator. 
The resulting algorithm falls in the \textit{noisy MCMC} framework \citep{alquier} as it simulates a chain that is not invariant for the target distribution.

As pointed out in \cite{alquier}, noisy MCMC is connected to pseudo-marginal algorithms \citep{andrieu}, where an unbiased and positive estimate of the target density is required.
In the presence of an intractable normalising constant, the pseudo-marginal approach requires an unbiased estimate of $1/z(\param)$. However, the reciprocal $1/\hat{z}(\param)$ yields a biased approximation of $1/z(\param)$ and so is not directly applicable to inference using pseudo-marginal techniques.
Recently, \cite{lyne} addressed this bias with Russian Roulette sampling, which yields an asymptotically exact algorithm. The implementation of the algorithm is computationally expensive, however, creating difficulties when inferring the model parameters for GRFs. 
Non-negative unbiased estimators have also been studied in \cite{jacob2015}, which showed that finding such an estimate is very challenging.

Consequently, we do not pursue such a computational approach in this paper. Instead, motivated by the inefficiency of a standard RJMCMC algorithm, we develop a noisy RJMCMC sampler that targets an approximated posterior distribution, rather than the desired one. We extend the theoretical analysis of noisy MCMC algorithms proposed in \cite{alquier} to trans-dimensional kernels, providing bounds on the total variation between the Markov chain of a noisy RJMCMC algorithm and a Markov chain with the desired target distribution under certain conditions.

We show that noisy RJMCMC algorithms are only useful when the estimator of the ratio of normalising constants has a small variance. Motivated by \cite{gelman_98}, we propose a smoother transition path between different models and resort to an alternative estimator with lower variance.
We demonstrate empirically that this idea simultaneously: (i) improves the mixing of the RJ Markov chain and (ii) decreases the asymptotic bias between the exact RJ Markov chain and its noisy approximation. Finally, we construct efficient jump proposal distributions for random walk (RW) noisy RJMCMC, which could be useful in the context of a large number of nested competing models.

The outline of the article is as follows. 
Section \ref{section:bmc} introduces the reader to basic concepts regarding Bayesian model comparison and to the reversible jump MCMC formulation.
In Section \ref{section:rjmcmc_intract} we discuss the extension of the reversible jump methodology to doubly-intractable posterior distributions and present theoretical properties and practical aspects of the RJMCMC samplers under such computational difficulties. 
Section \ref{section:rj_mcmc_samplers} presents some proposal tuning strategies for noisy RJMCMC.
In Section \ref{section:noisy_rjmcmc} we study the theoretical behavior of the noisy RJMCMC algorithm and derive convergence bounds.
We investigate the performance of noisy RJMCMC with a detailed numerical study that focuses on social network analysis in Section \ref{section:applications}.
We conclude the paper in Section \ref{section:discussion} with final remarks.

\section{Preliminaries}\label{section:bmc}
Suppose a finite set of competing models $\models=\{\models_1,\models_2,\models_3,\ldots\}$ are under consideration to describe the data $y$. In the Bayesian setting each model $\models_m$, where $m\in\{1,2,3,\ldots\}$, is characterised by a likelihood function $f(y\mid\param_m,\models_m)=f_m(y\mid\param_m)\propto q_m(y\mid\param_m)$, parameterised by an unknown parameter vector $\param_m\in\paramspa_m$. 
Each model is also associated with a prior distribution $p(\param_m\mid\models_m)=p_m(\param_m)$, used to express the beliefs about the parameter vector prior to observing the data $y$. 
The focus of interest in Bayesian inference for each competing model is the posterior distribution
\begin{equation}\label{eqn:postdistr1}
\pi(\param_m\mid y,\models_m)=
\frac{f_m(y\mid\param_m)p_m(\param_m)}{\pi(y\mid \models_m)}.
\end{equation}
The prior beliefs for each model are expressed through a prior distribution $p(\models_m)$, such that $\sum_{m\in\models}p(\models_m)=1$.

\subsection{Bayesian model comparison}
The marginal likelihood or model \textit{evidence} for model $\models_m$ is 
\begin{equation*}\label{eqn:evidence}
\pi(y\mid \models_m) = \int_{\paramspa_m} f_m(y\mid\param_m)p_m(\param_m)\,\mathrm{d}\param_m
\end{equation*}
and is rarely analytically tractable. 
However, knowledge of the evidence is required for a quantitative discrimination between competing models with the posterior model probabilities.
Using Bayes' theorem the posterior model probability for model $\models_m$ is
\begin{equation}\label{eqn:post_model_prob}
\pi(\models_m\mid y) = \frac{\pi(y\mid \models_m)p(\models_m)}{\sum_{j=1}^{|\models|} \pi(y\mid \models_j)p(\models_j)},
\end{equation}
where $|\models|$ is the cardinality of the model set. The probabilities $\pi(\models_m\mid y)$ are treated as a measure of the uncertainty of model $\models_m$.
Comparison of two competing models in the Bayesian setting is performed through the Bayes factor
\begin{equation*}\label{eqn:BF}
BF_{m,m'}=\frac{\pi(y\mid\models_m)}{\pi(y\mid\models_{m'})},
\end{equation*}
which provides evidence in favor of model $\models_m$ compared with model $\models_{m'}$. Using \eqref{eqn:post_model_prob}, the Bayes factor can also be expressed as the ratio of the posterior model odds to the prior odds,
\begin{equation*}\label{eqn:BF2}
BF_{m,m'}=\frac{\pi(\models_m\mid y)}{\pi(\models_{m'}\mid y)}/\frac{p(\models_m)}{p(\models_{m'})}.
\end{equation*}
A comprehensive review of Bayes factors is presented by \cite{kassr95}. 
There are at least two computational approaches to estimate Bayes factors: (i) within-model simulation, where the evidence is estimated separately for each model, see \cite{wyse_review} for a recent review on related methods; (ii) across-model simulation with trans-dimensional MCMC methods, which involves estimation of posterior model odds for chosen prior model odds. Reversible jump MCMC \citep{green95} is a popular trans-dimensional MCMC method and is the focus of this work.

\subsection{Reversible jump MCMC}\label{section:rjmcmc}
Reversible jump MCMC \citep{green95} generalises the Metropolis-Hastings (MH) algorithm \citep{metropolis1953equation,hastings} to allow for sampling from a distribution on a union of spaces of possibly differing dimensions, permitting state-dependent choices of move types. Let $\targ$ be a distribution over the general state space $\mathcal{X}$, 
\begin{equation*}
\mathcal{X}=\bigcup_{m\in\models}\left(\{m\},\paramspa_m\right),\qquad \models=\{\models_1,\models_2,\models_3,\ldots\},\quad \paramspa_m\subseteq\rset^{d_m}.
\end{equation*}
The dimension of the parameter space $\paramspa_m$ of model $\models_m$ is denoted by $d_m$.
The target $\targ$ is a joint posterior distribution on a model and a parameter,
\begin{linenomath}
\begin{equation}\label{eq:joint_posterior}
\pi(\models_m,\param_m\mid y)=\frac{f_m(y\mid\param_m)p_m(\param_m)p(\models_m)}{\sum_{k=1}^{|\models|}f_k(y\mid\param_k)p_k(\param_k)p(\models_k)\,\mathrm{d}\param_k}.
\end{equation}
\end{linenomath}
which can be factorised as the product of posterior model probabilities and model-specific parameter posteriors,
\begin{linenomath}
\begin{equation}\label{eq:joint_posterior_rj}
\pi(\models_m,\param_m\mid y)=\pi(\models_m\mid y)\pi(\param_m\mid y,\models_m).
\end{equation}
\end{linenomath}
To sample the model indicator and the model parameters jointly a Markov chain is constructed with state space $\mathcal{X}$ and stationary distribution $\pi(\models_m,\param_m\mid y)$. The state space $\mathcal{X}$ is a finite union of subspaces of possibly varying dimensions. By marginalisation, we obtain the probability of being in subspace $\paramspa_m$.

The reversible jump MCMC scheme allows for Metropolis-Hastings moves between states defined by $x=(m, \param)$ and $x'=(m', \param')$ that may have different dimensions $d_m$ and $d_{m'}$, respectively. Since it is a MH-type algorithm it is $\pi$-reversible and thus $\pi$-invariant. As a consequence, the Markov chain simulated by RJMCMC produces samples $(m,\param_m)\sim \pi(\cdot\mid y)$.

Below we formulate the RJMCMC algorithm in a way that generalises non trans-dimensional MCMC. This will ease the analysis of Section \ref{section:noisy_rjmcmc}.
Let $h(\param',m'\mid\param,m)=\omega(m,m')T_{m,m'}(\param, \param')$ be the proposal distribution for the transition from 
$(m,\param_m)$ to $(m',\param_{m'})$. We denote by $\omega(m,m')$ the probability of proposing a jump to model $\models_{m'}$ when the chain is currently at model $\models_{m}$ and by $T_{m,m'}(\param, \param')$ the proposal for the parameter vector. 
For simplicity, we consider the case of RW updates when there are common parameters between the current and the proposed states.
We define $\mathcal{C}_{m,m'}$ as the indicator of the common parameters between models $\models_{m}$, $\models_{m'}$ 
so that $\param_{\mathcal{C}_{m,m'}}$ is a sub-vector of model parameters and by $\overline{\mathcal{C}}_{m',m}$ the set of parameters that are in $\models_{m'}$, but not in $\models_{m}$. Then the proposed parameter vector is defined as $\param':=[\param_{\mathcal{C}_{m,m'}},\param'_{\overline{\mathcal{C}}_{m',m}}]\in\rset^{d_{m'}}$. and the proposal distribution for the parameter vector takes the form
\begin{equation}
T_{m,m'}(\param,\param')=
\begin{dcases}
g(\param'_{\overline{\mathcal{C}}_{m',m}}\mid \param_{\mathcal{C}_{m,m'}})& \text{if } m\neq m', \\
w(\param'\mid\param) & \text{if } m=m', 
\end{dcases}
\end{equation}
where $g(\cdot\mid\param)$ and $w(\cdot\mid\param)$ are some proposal distributions.
The RJMCMC scheme proceeds as in Algorithm \ref{alg:rjmcmc}. \cite{green95,green:book} offers a representation in terms of random numbers and a Jacobian term to eliminate the apparent difficulty of moving between spaces of different dimensions.

The advantage of RJMCMC is that estimates of the posterior model probabilities are readily available along with parameter estimates for each competing model. Of course, if the Markov chain never visits a model because it is unlikely a-posteriori then the parameter estimates will not be available for that model.
To assess support for the models under examination with the RJMCMC scheme, the output from the trans-dimensional chain is processed to calculate the Bayes factor. 
Assuming equal prior probabilities on models $\models_{m}$ and $\models_{m'}$, this motivates the simple estimate of $B_{m,m'}$ as $F_{m}/F_{m'}$, where $F_{m}$ is the frequency of visits (out of a chain of length $G$) to model $\models_{m}$. 
\begin{algorithm}[H]
\caption{Reversible jump MCMC}
\label{alg:rjmcmc}
\begin{algorithmic}[1]
\State Initialise $(m_0,\param_0)$.
  \For {$n~=~0,\ldots,G-1$}
	  \State Select a candidate model $\models_{m'}$ with probability $\omega(m_{(n)},m')$.
		\State Propose $\param'\sim T_{m_{(n)},m'}(\param_{(n)}, \param')$.
    \LongState{Estimate the acceptance probability $A_{m_{(n)},m'}(\param_{(n)},\param')=\min\{1,\rho_{m_{(n)},m'}(\param_{(n)},\param')\}$,
\begin{equation}\label{eq:rjmcmc_ratio}
\rho_{m_{(n)},m'}(\param_{(n)},\param')=
\frac{f_{m'}(y\mid\param')}{f_{m_{(n)}}(y\mid\param_{(n)})}
\frac{p_{m'}(\param')}{p_{m_{(n)}}(\param_{(n)})}
\frac{p(\models_{m'})}{p(\models_m)}
\frac{\omega(m',m_{(n)})}{\omega(m_{(n)},m')}
\frac{T_{m',m_{(n)}}(\param', \param_{(n)})}{T_{m_{(n)},m'}(\param_{(n)}, \param')}.
\end{equation}
}
\vspace{-1.4em}
	\State Set $(m_{(n+1)},\param_{(n+1)})~\leftarrow~(m',\param^{\prime})$ with probability $A$, else $(m_{(n+1)},\param_{(n+1)})~\leftarrow~(m_{(n)},\param_{(n)})$.
 \EndFor
\State {\bf{return}} $\{m_n,\param_n\}_{n=1,\ldots,G}$
\end{algorithmic}
\end{algorithm}

\section{Reversible jump MCMC in the presence of likelihood intractability}\label{section:rjmcmc_intract}
The reversible jump algorithm presents an asymptotically (as the number of transitions $G\rightarrow\infty$) exact MCMC method by targeting the posterior distribution of interest, $\pi$. When the posterior involves an intractable likelihood a naive implementation of RJMCMC for GRFs is challenging because it is simply not possible to simulate a transition of this exact chain.

Indeed, given the current state $(m,\param)$, the proposed value of the next state of the chain, $(m',\param')$, is accepted with probability $A_{m,m'}(\param,\param')=\min\{1,\rho_{m,m'}(\param,\param')\}$, where
\begin{align} \label{eqn:MH_accept_prob}
\rho_{m,m'}(\param,\param')=
\frac{q_{m'}(y\mid\param')}{q_m(y\mid\param)}
\frac{p_{m'}(\param')}{p_{m}(\param)}
\frac{p(\models_{m'})}{p(\models_m)}
\frac{\omega(m',m)}{\omega(m,m')}
\frac{T_{m',m}(\param', \param)}{T_{m,m'}(\param, \param')}
\times
\frac{z_m(\param)}{z_{m'}(\param')}.
\end{align}
The acceptance probability is dependent on a ratio of intractable normalising constants and, thus, cannot be directly evaluated. GRF models belong to the exponential family and satisfy the identity
\begin{equation}\label{eqn:unbiased_IS}
\frac{z_m(\param)}{z_{m'}(\param')}
= 
\sum_{y\in \mathcal{Y}}^{}\frac{q_m(y\mid \param)}{q_{m'}(y\mid \param')}\frac{q_{m'}(y\mid \param')}{z_{m'}(\param')}
= 
\esp_{y\sim f_{m'}(\cdot\mid\param')}\left[\mathpzc{w}(y)\right].
\end{equation}
This expectation can be estimated with importance sampling, where $\mathpzc{w}(y)=q_m(y\mid\param)/q_{m'}(y\mid\param')$ are the importance sampling weights. 

Below we present variants of the RJMCMC algorithm that bypass the need to calculate the intractable ratio $z_m(\param)/z_{m'}(\param')$ in \eqref{eqn:MH_accept_prob}, by replacing it with an unbiased estimator. 
It is important to note that using an estimator instead of the true ratio has significant consequences on the convergence of the Markov chain. 
In particular, the Markov chain is usually no longer asymptotically exact and in this context, such an algorithm is referred to as \textit{noisy} MCMC.
However, \cite{alquier} have shown that the distribution of the noisy MCMC Markov chain can be made arbitrarily close to the desired stationary distribution, provided that an estimator with an arbitrarily small variance is available.
In the following, we refer to as noisy RJMCMC the implementation of the RJMCMC algorithm where the ratio $z_m(\param)/z_{m'}(\param')$ is estimated. 

The efficiency of the noisy RJMCMC depends on the asymptotic approximation of $\pi$, an estimator of $z_m(\param)/z_{m'}(\param')$ with small variance and the mixing property of the chain.
The first and second points are discussed in the remainder of this Section through the presentation of different estimators of the ratio of normalising constants. 
The third point is considered at Section \ref{section:rj_mcmc_samplers}, where efficient proposal distributions are designed.

\subsection{Reversible jump exchange algorithm}
\cite{murray} presented the exchange algorithm that allows inference for doubly-intractable distributions and circumvents the issue of intractability of \eqref{intro_eqn:likelihood}. 
The exchange algorithm introduces an auxiliary variable $y'\sim f_{m'}(\cdot\mid \param')$ that is used to estimate the intractable ratio of normalising constants with a one-sample unbiased importance sampling estimator of \eqref{eqn:unbiased_IS}.

Pseudo-marginal algorithms \citep{andrieu} share the same principles with the exchange algorithm, replacing the intractable likelihood function with a positive and unbiased estimate that can be obtained using likelihood simulations or importance sampling. Like the exchange algorithm, pseudo-marginal MCMC defines an asymptotically exact MCMC algorithm.

\cite{caimo2} devised a trans-dimensional extension of the exchange algorithm which is $\pi$-invariant and generalises the exchange algorithm of \cite{murray} to trans-dimensional settings in the same way that RJMCMC generalises the Metropolis-Hastings algorithm. 
We refer to this instance as the \textit{RJ exchange} when $N=1$ draws from the likelihood are used to estimate \eqref{eqn:unbiased_IS}.

Despite yielding a Markov chain that is $\pi$-invariant, the one-sample unbiased importance sampling estimator usually has a large variance (see discussion below). Empirical results in \cite{alquier} suggest that such an exchange algorithm is likely to be inefficient because of slow mixing properties and propose taking multiple auxiliary draws $N>1$. 
In the trans-dimensional setting, this motivates a noisy RJMCMC algorithm that uses $N>1$ likelihood draws or another estimator with small variance.

\begin{figure}[H]
\centering
\includegraphics[width=16cm,height=11cm]{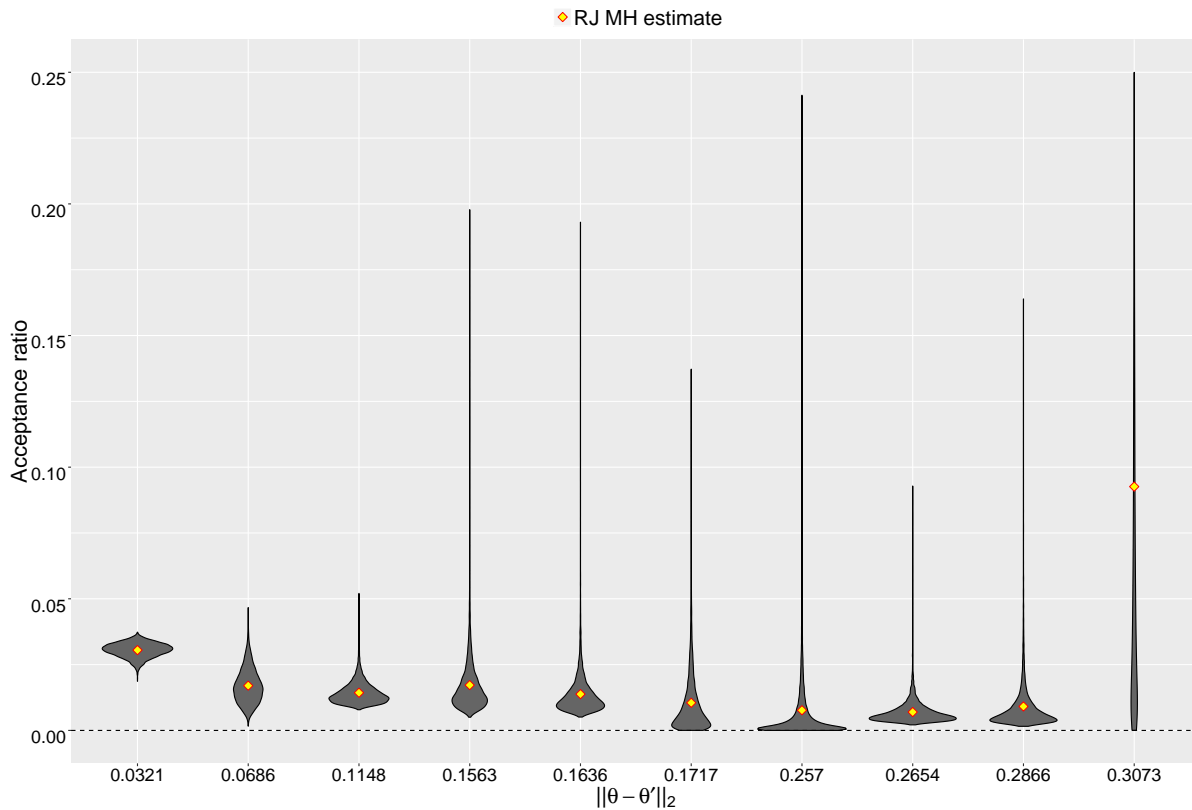}
\caption{In the context of a trans-dimensional move between 2 different exponential random graph (ERG) models (see Section \ref{section:applications}), 10 samples $(\param,\param')$ are drawn at random. 
For each of them we plot the MH acceptance rate (yellow spade) and the distribution of the exchange acceptance rate (violin plot). 
In practice the MH acceptance rate is not tractable but we use \eqref{eqn:biased_IS} with $N=5\times 10^5$ to estimate it precisely.
As the distance between $\param$ and $\param'$ increases, the acceptance ratio in the RJ exchange algorithm becomes more variable. 
} 
\label{fig:noisy_RJ_motivation}
\end{figure}

A graphical illustration that attempts to explain the inefficiency of the RJ exchange can be seen in Figure \ref{fig:noisy_RJ_motivation}. 
This example involves transitions with a random walk proposal from ERG model $\models_1$ to model $\models_2$, which we provide details of in Section \ref{section:karate}. 
It is evident that the RJ exchange acceptance ratio will frequently underestimate the RJ MH acceptance ratio, making it more likely to reject the proposed move and affecting the convergence speed of the algorithm, particularly as the $L^2$ norm $\left\Vert \param-\param' \right\Vert_2$ increases. 

\subsection{Noisy reversible jump MCMC}\label{section:ration_est}
Estimating ratios of intractable normalising constants using \eqref{eqn:unbiased_IS} has been proposed in \cite{alquier}. 
A Monte Carlo estimator based on multiple auxiliary draws from the likelihood with respect to the proposed parameter, $y'_1, y'_2,\ldots,y'_{N}\sim f_{m'}(\cdot\mid\param')$, approximates \eqref{eqn:unbiased_IS} by
\begin{align}\label{eqn:biased_IS}
\widehat{\frac{z_m(\param)}{z_{m'}(\param')}}=\frac{1}{N}\sum_{i=1}^{N}\frac{q_m(y'_i\mid\param)}{q_{m'}(y'_i\mid\param')}.
\end{align}
For the remainder of the paper, we will refer to \eqref{eqn:biased_IS} as the standard importance sampling estimator (ISE). The special case where $N=1$ yields the RJ exchange algorithm. We now address the issue of estimator variance.

The work of \cite{stoehr_hmc} that considered non trans-dimensional moves demonstrated that the quality of the ISE strongly decreases when the distance $\left\Vert \param-\param' \right\Vert_2$ increases. This is even more involved when $\param$ and $\param'$ belong to different parameter spaces. 
\cite{boland} proved that the variance of the estimator \eqref{eqn:biased_IS} is controlled when the distance between the parameters decreases. 
This result can be extended to jumps between models of different dimensions.
For instance, in the case of an independence sampler the full parameter vector is updated. 
Consequently, $\left\Vert \param-\param' \right\Vert_2$ can be large and so may be the variance of the estimator. This is a serious concern that may affect the performance of the algorithm, as mentioned at the beginning of Section \ref{section:rjmcmc_intract}.

To study the variance of the ISE in the trans-dimensional setting, we consider for simplicity the case of a random walk transition between two nested models where $d_m<d_{m'}$ and define $\dot{\param}:=[\param_{\mathcal{C}_{m,m'}},0_{d_{m'}-d^{}_m}]\in\rset^{d_{m'}}$.
A RW allows us to control the distance $\left\Vert \dot{\param}-\param' \right\Vert_2$, as it is simply equal to $\left\Vert \param'_{\overline{\mathcal{C}}_{m',m}} \right\Vert_2$, and to apply Proposition \ref{prop:var_bound} below.
\begin{proposition}[Proposition 1 in \cite{boland}]
\label{prop:var_bound}
For any GRF model and $(\dot{\param},\param')\in \paramspa_{m'} \times \paramspa_{m'}$, the variance (denoted by the variance operator $\var$) of the estimator in \eqref{eqn:biased_IS} decreases when $\left\Vert \dot{\param}-\param' \right\Vert_2 \downarrow 0$, such that
\begin{equation}\label{eq:noisy_rj_est_ratio}
\var_{y'_1, y'_2,\ldots,y'_{N}\sim f_{m'}(\cdot\mid \param')}\left[\widehat{\frac{z_{m'}(\dot{\param})}{z_{m'}(\param')}}\right]=\mathcal{O}(\left\Vert \dot{\param}-\param' \right\Vert_2),
\end{equation}
\end{proposition}
To further decrease the variance in \eqref{eq:noisy_rj_est_ratio} we consider an alternative estimator that takes advantage of the auxiliary draws from the likelihood, and one which has also been considered in previous works \citep{friel4,bouranis2,stoehr_hmc,Everitt2}. Let us introduce an auxiliary variable $t\in[0,1]$ and discretise $[0,1]$ as $0=t^{}_1<\ldots<t^{}_L=1$, where $L\in\mathbb{Z}_{+}$. 
We remain in the case of nested model jumps and define the vector-valued mapping 
\begin{equation*}
\param'(t):=
\begin{dcases}
(1-t)\dot{\param}+t\param'  & \text{if } d_m<d_{m'}, \\
(1-t)\param+t\param'        & \text{if } d_m=d_{m'}. \\
\end{dcases}
\end{equation*}
The idea is to replace the ratio of normalising constants by a telescopic product of ratios of normalising constants taken at two consecutive points of the path $t\in[0,1]$, so that $(\param'(t^{}_j),\param'(t^{}_{j+1}))$ are close to each other. More precisely, the ratio of normalising constants can be written as 
\begin{equation}\label{eqn:PSE_ratio1}
\frac{z_{m'}(\param'(t^{}_1))}{z_{m'}(\param'(t^{}_L))}
=
\prod_{j=1}^{L-1}\frac{z_{m'}(\param'(t^{}_j))}{z_{m'}(\param'(t^{}_{j+1}))}.
\end{equation}
Any factor in the RHS of this equation is estimated as follows.
The un-normalised likelihood $q_{m'}(y\mid \param'(t^{}_j))$ can be considered as an importance distribution for the "target" distribution $f_{m'}(y\mid \param'(t^{}_j))$, noting that 
\begin{equation*}
\frac{z_{m'}(\param'(t^{}_j))}{z_{m'}(\param'(t^{}_{j+1}))}=\esp_{y^{(j+1)}\sim f_{m'}(\cdot\mid\param'(t^{}_{j+1}))}
\left[
\frac{q_{m'}(y^{(j+1)}\mid \param'(t^{}_j))}{q_{m'}(y^{(j+1)}\mid \param'(t^{}_{j+1}))}
\right].
\end{equation*}
An unbiased importance sampling estimator of this expectation can be obtained by simulating multiple draws $y^{(j+1)'}_1,\dots,y^{(j+1)'}_S \sim f_{m'}(\cdot\mid \param'(t^{}_{j+1}))$, yielding
\begin{equation*}
\widehat{\frac{z_{m'}(\param'(t^{}_j))}{z_{m'}(\param'(t^{}_{j+1}))}}=
\frac{1}{S}\sum_{b=1}^{S}\frac{q_{m'}(y^{(j+1)'}_b\mid \param'(t^{}_j))}{q_{m'}(y^{(j+1)'}_b\mid \param'(t^{}_{j+1}))}.
\end{equation*}
Since the expectation of a product of independent random variables is the product of the expectations, then 
\begin{equation}\label{eqn:PSE_ratio}
\widehat{\frac{z_{m'}(\param'(t^{}_1))}{z_{m'}(\param'(t^{}_L))}}
:=
\prod_{j=1}^{L-1}\widehat{\frac{z_{m'}(\param'(t^{}_j))}{z_{m'}(\param'(t^{}_{j+1}))}}
=
\prod_{j=1}^{L-1}\left[\frac{1}{S}\sum_{b=1}^{S}\frac{q_{m'}(y^{(j+1)'}_b\mid \param'(t^{}_j)))}{q_{m'}(y^{(j+1)'}_b\mid \param'(t^{}_{j+1}))}\right]
\end{equation}
gives an unbiased estimator of the intractable ratio of normalising constants. 

This telescopic product estimator (TPE) will come at an extra computational cost, similarly to \eqref{eqn:biased_IS}. The larger the number of path steps $L$ and the number of draws $S$ (where $N=(L-1)\times S$), the more precise the estimate of the ratio of normalising constants at the expense of computational time. Nevertheless, empirical results in Section \ref{section:applications} show that this estimator can lead to improved performance results compared to the ISE for the same number of likelihood draws in the estimator. 
Algorithm \ref{alg:_noisy_rjmcmc} contains pseudo-code for the noisy RJMCMC algorithm.
\begin{algorithm}[H]
\caption{Noisy reversible jump MCMC}
\label{alg:_noisy_rjmcmc}
\begin{algorithmic}[1]
\State Initialise $(m_0,\param_0)$.
  \For {$n~=~0,\ldots,G-1$}
	  \State Select a candidate model $\models_{m'}$ with probability $\omega(m_{(n)},m')$. 
		\State Propose $\param'\sim T_{m_{(n)},m'}(\param_{(n)}, \param')$.
		\LongState{Draw $y'_1,\ldots,y'_N \sim f_{m'}(\cdot\mid \param')$ depending on the type of estimator of the ratio $z_{m_{(n)}}(\param_{(n)})/z_{m'}(\param')$, using \eqref{eqn:biased_IS} or \eqref{eqn:PSE_ratio}.}
    \LongState{Estimate the acceptance probability $\hat{A}_{m_{(n)},m'}(\param_{(n)},\param',y')=\min\{1,\hat{\rho}_{m_{(n)},m'}(\param_{(n)},\param',y')\}$,
\begin{align}\label{eq1}
\hat{\rho}_{m_{(n)},m'}(\param_{(n)},\param',y')
&=
\frac{q_{m'}(y\mid\param')}{q_{m_{(n)}}(y\mid\param_{(n)})}
\frac{p_{m'}(\param')}{p_{m_{(n)}}(\param_{(n)})}
\frac{p(\models_{m'})}{p(\models_{m_{(n)}})}
\frac{\omega(m',m_{(n)})}{\omega(m_{(n)},m')}
\frac{T_{m',m_{(n)}}(\param', \param_{(n)})}{T_{m_{(n)},m'}(\param_{(n)}, \param')} \nonumber\\
& \mathrel{\phantom{=}}
\times
\widehat{\frac{z_{m_{(n)}}(\param_{(n)})}{z_{m'}(\param')}}.
\end{align}
}%
\vspace{-1.4em}
	\State Set $(m_{(n+1)},\param_{(n+1)})~\leftarrow~(m',\param^{\prime})$ with probability $\hat{A}$, else $(m_{(n+1)},\param_{(n+1)})~\leftarrow~(m_{(n)},\param_{(n)})$.
 \EndFor
\State {\bf{return}} $\{m_n,\param_n\}_{n=1,\ldots,G}$
\end{algorithmic}
\end{algorithm}
It is worth noting that \cite{Karagiannis} use a similar telescopic product, where the transition path is used to introduce a trans-dimensional Annealed importance sampling (AIS) estimator of the ratio $\pi(m',\theta')/\pi(m,\theta)$ that appears in \eqref{eq:rjmcmc_ratio}. 
Contrary to the noisy RJMCMC algorithm which uses an importance sampling estimator of the ratio of likelihood normalising constants, the proposed RJMCMC algorithm of \cite{Karagiannis} does not target a similar posterior distribution as the noisy RJMCMC algorithm and the trans-dimensional AIS estimator is constructed in a different way.

\section{Choice of reversible jump proposal densities}\label{section:rj_mcmc_samplers}
All algorithms in this paper were run using a proposal distribution $\omega(m,m')$ for the transition from $(m,\param_m)$ to $(m',\param_{m'})$ that was set to a discrete Uniform $\mathcal{U}(1,|\models|)$.
Below we present the strategies that we followed in order to tune the proposal distribution for the parameter vector.

\subsection{Independent proposals}
We adopted the auto-reversible jump (Auto-RJ) exchange algorithm that was proposed by \cite{caimo3} to perform Bayesian model comparison of exponential random graph models in the area of Bayesian network analysis.

The Auto-RJ takes the form of an independence sampler, making use of a parametric approximation of the posterior distribution for each model in an attempt to bypass the issue of tuning the parameters of the jump proposal distributions and increase within-model acceptance rates.
The Auto-RJ consists of pilot MCMC runs that are used to sample from the posterior distribution of each competing model with the exchange algorithm \citep{murray} and then to approximate the estimated posterior by Gaussian distributions determined by the first moments of each sample.
The online step of the Auto-RJ sets $T_{m,m'}(\param, \param')$ to this approximation of the posterior density for both the within- and between-model jumps.

The TPE in \eqref{eqn:PSE_ratio} is particularly suitable to the independence sampler, as it allows for "long" transitions (in terms of a larger $L^2$ norm in this context) while having an estimator of the intractable ratio of normalising constants with small variance, compared to the ISE. 
On the other hand, the requirement of pilot MCMC runs reduces the appeal of this method when the number of models under investigation is large.
Indeed, in such a case a within-model simulation approach may be as efficient. 
Additionally, the Auto-RJ is likely to fail when the posterior of at least one model is multi-modal.

\subsection{Random walk proposals}
A natural alternative to the independence sampler comes from allowing the proposal $(m',\param')$ to depend on the current state $(m,\param)$. 
Despite eliminating the need for pilot runs, an open question remains as to how optimally estimate the proposal variance parameter. 
An inefficient proposal mechanism will result in a Markov chain which slowly explores the state space and has high auto-correlation, increasing the asymptotic variance of Monte Carlo estimators. 
Such inefficiency can be caused by not proposing large moves away from the current state of the chain or by proposing moves with prohibitively small associated acceptance probabilities. 

For the within-model Gaussian RW updates (see below), a proposal distribution with a variance-covariance matrix in the form of $\Sigma_{m'} = \lambda_{m'}(\Omega_{m'} + C^{-1}_{m'})^{-1}$ was assumed, to account for possible correlations between the model parameters \citep{chib2,mcmcpack}. 
The prior precision is denoted by $\Omega_{m'}$ and $\lambda_{m'}\in\rset^{+}$ is a positive Metropolis tuning parameter. 
The precision matrix $C^{-1}_{m'}$ is set as the negative Hessian $-\nabla^{2}_{\param_{m'}}\log{f_{m'}(y\mid\param_{m'})}|_{\htheta^{MLE}_{m'}}$. 
Estimation of the MLE for each ERG model in Section \ref{section:applications} was performed with the Monte Carlo Maximum Likelihood Estimation (MC-MLE) procedure proposed by \cite{geyer}. 
Alternative procedures exist, see \cite{hunter}.
The gradient of the log-likelihood function can be written as
\begin{align}\label{eq:grf_grad}
\nabla_{\param_{m'}}\log{f_{m'}(y\mid\param_{m'})}
&=s_{m'}(y)-\frac{\nabla_{\param_{m'}}z_{m'}(\param_{m'})}{z_{m'}(\param_{m'})} \nonumber\\
&=s_{m'}(y)-\frac{\sum_{y\in \mathcal{Y}}^{} s_{m'}(y)\text{exp}\left\{\param^{\top}_{m'} s_{m'}(y)\right\}}{\sum_{y\in \mathcal{Y}}^{}\text{exp}\left\{\param^{\top}_{m'} s_{m'}(y)\right\}} \nonumber\\
&=s_{m'}(y)-\mathbb{E}_{y\mid \param_{m'}}\left[s_{m'}(y)\right].
\end{align}
Then the Hessian matrix can be found using the identity
\begin{align}\label{eq:grf_hessian}
\nabla^{2}_{\param_{m'}}\log{f_{m'}(y\mid\param_{m'})}
&=\nabla_{\param_{m'}}\left[-\frac{\nabla_{\param_{m'}}z_{m'}(\param_{m'})}{z_{m'}(\param_{m'})}\right]\nonumber\\
&=-\bigg\{\mathbb{E}_{y\mid \param_{m'}}\left[s^2_{m'}(y)\right]-\left[\mathbb{E}_{y\mid \param_{m'}}\left[s_{m'}(y)\right]\right]^2\bigg\}\nonumber\\
&=-\mathbb{V}_{y\mid \param_{m'}}\left[s_{m'}(y)\right],
\end{align}
where $\mathbb{V}_{y\mid \param_{m'}}\left[s_{m'}(y)\right]$ denotes the covariance matrix of the vector $s_{m'}(y)$ with respect to $f_{m'}(y\mid\param_{m'})$. 
For GRF models those Hessian matrices are intractable and so we resort to Monte Carlo sampling from $f_{m'}(y\mid\param_{m'})$ in order to estimate $\mathbb{V}_{y\mid \param_{m'}}\left[s_{m'}(y)\right]$. This setting helped us reach a reasonable mixing rate within each model.

For the between-model moves, a popular choice in the implementation of the RJMCMC algorithm for nested cases is the second order method of \cite{brooks}. 
The second order method is based on a Taylor series expansion of the acceptance probability around certain canonical jumps, providing a fully automated framework for achieving local adaptation of the proposal density. 
It attempts to maximise the marginal acceptance probability $A_{m,m'}(\param,\param)$ and not the extended acceptance probability $A_{m,m'}\left(\param,\param,y'\right)$ that uses the auxiliary data $y'$ and so it is incompatible with the reversible jump exchange algorithm.
While such a strategy would be sensible for noisy RJMCMC, it involves estimation of \eqref{eq:grf_grad} and \eqref{eq:grf_hessian}, which are unavailable in closed form for models like GRFs.
The gradients can be numerically approximated with Monte Carlo simulation; this relies on repeated likelihood simulations, increasing considerably the computational cost per iteration of the noisy RJMCMC algorithm and so we do not pursue this strategy further.

Our implementation of the between-model moves is inspired by \cite{ehlers2}, who proved that the full-conditional posterior distribution, $\pi(\param'_{\overline{\mathcal{C}}_{m',m}}\mid y,\param_m,\models_{m'})$, which is based upon the current state of the chain, is the optimal proposal distribution for the random vector $\param'_{\overline{\mathcal{C}}_{m',m}}$.
We tackled the intractability of the full-conditional by first considering a multivariate Gaussian distribution $\mathcal{MVN}\left(\htheta^{MLE}_{m'},\Sigma_{m'}\right)$, assuming diffuse prior distributions.
Standard theory can be used to derive a mean $\mu^*$ and covariance matrix 
$\Sigma^*_{m'}$ and approximate the full-conditional distribution with a multivariate Gaussian distribution, $\mathcal{MVN}\left(\param'_{\overline{\mathcal{C}}_{m',m}};\mu^*,\Sigma^*_{m'}\right)$. 
To summarise, the proposal distribution for our implementation of the noisy Gaussian RW RJMCMC is
\begin{equation}
T_{m,m'}(\param, \param')=
\begin{dcases}
\mathcal{MVN}\left(\param'_{\overline{\mathcal{C}}_{m',m}};\mu^*,\Sigma^*_{m'}\right)  & \text{if } m\neq m', \\
\mathcal{MVN}\left(\param';\param,\Sigma_{m'}\right) & \text{if } m=m'. 
\end{dcases}
\end{equation}
Finally, \cite{Karagiannis} construct efficient RJMCMC transitions, relying on the assumption that the target distribution $\pi(\models_m,\param_m\mid y)$ is known up to a normalising constant. For the reasons explained at Section \ref{section:ration_est}, we do not pursue that methodology further.

\section{Theoretical guarantees for noisy reversible jump MCMC}\label{section:noisy_rjmcmc} 
The goal of this section is to extend the results of \cite{alquier} to the trans-dimensional setting and then to apply these to the context of noisy RJMCMC 
with the different estimators developed earlier. To do so, we first present RJMCMC as a simple generalisation of the MH algorithm. Following Section \ref{section:rjmcmc}, we consider again the general state space $\mathcal{X}$ that defines the target distribution $\pi$. Our interest lies in constructing a Markov chain with transition kernel $\mathcal{P}$, with $\pi$ as the invariant distribution.

Let us consider $\pi$ as a probability measure on the compact set $\mathcal{X}$. Then $\pi$ is an invariant distribution if 
\begin{linenomath}
\begin{equation*}
\int_{\mathcal{X}}\pi(dx)\mathcal{P}(x,dx')=\pi(dx).
\end{equation*}
\end{linenomath}
A sufficient, but not necessary, condition is that the respective Markov chain is reversible, so that for all Borel sets $\mathcal{B,B'}\subset\mathcal{X}$
\begin{linenomath}
\begin{equation}\label{eq:det_bal_rjmcmc}
\int_{\mathcal{B}}\pi(dx)\mathcal{P}(x,B')=\int_{\mathcal{B}'}\pi(dx')\mathcal{P}(x',B).
\end{equation}
\end{linenomath}
Let $\upsilon(x,dx')$ be a proposal measure for the move from the current state $x=(m, \param)$ to the proposed state $x'=(m', \param')$, where $x'$ is accepted with probability $\alpha(x,x')=A_{m,m'}(\param,\param')$. 
The transition kernel is given by
\begin{linenomath}
\begin{equation}\label{eq:rjmcmc_kernel}
\mathcal{P}(x,B')=\int_{\mathcal{B'}}\upsilon(x,dx')\alpha(x,x')+\mathcal{R}(x)\mathbbm{1}(x\in\mathcal{B'}),
\end{equation}
\end{linenomath}
where
\begin{linenomath}
\begin{equation*}
\mathcal{R}(x)=\int_{\mathcal{X}}\upsilon(x,dx')[1-\alpha(x,x')]
\end{equation*}
\end{linenomath}
is the probability of rejecting any proposed move, while at state $x$. By substituting \eqref{eq:rjmcmc_kernel} into \eqref{eq:det_bal_rjmcmc}, it is straightforward to show that
\begin{linenomath}
\begin{equation*}
\int_{(x,x')\in\mathcal{B,B'}}\pi(dx)\upsilon(x,dx')\alpha(x,x')=\int_{(x,x')\in\mathcal{B,B'}}\pi(dx')\upsilon(x',dx)\alpha(x',x).
\end{equation*}
\end{linenomath}
The above formulation encompasses standard MCMC.

We denote by $\hat{\mathcal{P}}$ the transition kernel of the Markov chain resulting from the estimators \eqref{eqn:biased_IS} and \eqref{eqn:PSE_ratio}, where the acceptance ratio $\rho(x,x')$ in $\alpha(x,x')=\min\{1,\rho(x,x')\}$ is replaced by an estimator $\hat{\rho}(x,x',y')$, yielding $\hat{\alpha}(x,x',y')=\min\{1,\hat{\rho}(x,x',y')\}$. 
\begin{theorem}[Corollary 3.1 in \cite{mitrophanov}]\label{theorem1}
We assume that
\begin{description}
\item[(H1)] The reversible jump MH Markov chain with transition kernel $\mathcal{P}$ and acceptance probability $\alpha$ is uniformly ergodic, for which it holds that
\begin{linenomath}
\begin{equation*}
\sup_{x_0\in\mathcal{X}}\|\delta_{x_0}\mathcal{P}^n-\pi\|\leq Q\xi^n
\end{equation*}
\end{linenomath}
for some constants $Q<\infty$ and $\xi < 1$, where $\|\cdot\|$ is the total variation distance and $\delta$ is the Dirac delta function. 
\end{description}
Then for any transition $n\in\mathbb{N}$ and for any starting point $x_0\in\mathcal{X}$ it holds that
\begin{linenomath}
\begin{equation*}
\|\delta_{x_0}\mathcal{P}^n - \delta_{x_0}\hat{\mathcal{P}}^n\|
\leq 
\left(\phi + \frac{Q\xi^{\phi}}{1-\xi}\right)\|\mathcal{P}-\hat{\mathcal{P}}\|,
\end{equation*}
\end{linenomath}
where $\|\mathcal{P}-\hat{\mathcal{P}}\|=\sup_{x\in\mathcal{X}}\|\delta_x\mathcal{P}-\delta_x\hat{\mathcal{P}}\|$ is the total variation measure between the two kernels and $\phi=\left\lceil \frac{\log(1/Q)}{\log(\xi)} \right\rceil \leq n-1$.
\end{theorem}
An application of Theorem \ref{theorem1} is now provided when an approximation to the true transition kernel arises from Algorithm \ref{alg:_noisy_rjmcmc}.
\begin{corollary}\label{cor:noisy_rjmcmc_bound}
Let us assume that
\begin{description}
\item[(H1)] The Markov chain with transition kernel $P$ is uniformly ergodic holds.
\item[(H2)] There exists a function $\gamma:\mathcal{X}^2\to\rset^{+}$, such that for all $(x,x')\in \mathcal{X}^2$,
\begin{equation}\label{hyp_deviation}
\mathbb{E}_{y'\sim f(\cdot\mid x')}
\left|\hat{\rho}(x,x',y')-\rho(x,x')\right|
\leq \gamma(x,x').
\end{equation}
\end{description}
Then for any $n\in\mathbb{N}$ and for any starting point $x_0\in\mathcal{X}$ it holds that
\begin{equation*}
\|\delta_{x_0}\mathcal{P}^n - \delta_{x_0}\hat{\mathcal{P}}^n\|
\leq 
\left(\phi + \frac{Q\xi^{\phi}}{1-\xi}\right)\sup_{x}\int{\rm d}x' \upsilon(x,x') \gamma(x,x'),
\end{equation*}
where $\phi=\left\lceil \frac{\log(1/Q)}{\log(\xi)} \right\rceil$.
\end{corollary}
When the upper bound in \eqref{hyp_deviation} is uniformly bounded such that for all $(x,x')\in \mathcal{X}^2$ it holds that $\gamma(x,x')\leq \gamma <+\infty$, then
\begin{linenomath}
\begin{equation*}
\|\delta_{x_0}\mathcal{P}^n - \delta_{x_0}\hat{\mathcal{P}}^n\|
\leq 
\gamma\left(\phi + \frac{Q\xi^{\phi}}{1-\xi}\right).
\end{equation*}
\end{linenomath}
Consequently, letting $n\rightarrow\infty$ yields
\begin{linenomath}
\begin{equation*}
\limsup_{n\rightarrow\infty} \|\delta_{x_0} \hat{P}^n-\pi\|
\leq 
\gamma \left(\phi + \frac{Q\xi^{\phi}}{1-\xi}\right).
\end{equation*}
\end{linenomath}
Below we make a series of assumptions that will help us show that the noisy RJMCMC algorithm will yield a Markov chain which will converge to the target posterior density as: (i) $N\rightarrow \infty$, if the ratio $z_m(\param)/z_{m'}(\param')$ is estimated with \eqref{eqn:biased_IS} or (ii) $L,S\rightarrow \infty$, if the TPE in \eqref{eqn:PSE_ratio} is considered, instead. 
\begin{description}
\item[(A1)]For each prior over the model-specific parameters, there is a constant $c^{}_{p_m}$ such that \\$c^{-1}_{p_m}\leq p_m(\param)\leq c^{}_{p_m}\,,\forall m$.
\item[(A2)]There is a constant $c^{}_{\upsilon(x,x')}$ such that $c^{-1}_{\upsilon(x,x')}\leq \upsilon(x,x')=h(\param',m'\mid \param,m)\leq c_{\upsilon(x,x')}\,,\forall m,m'$.
\item[(A3)]A prior probability $0< p(\models_m)\leq c^{}_{\models_m}\leq 1\,,\forall m$ is assigned to each model.
\item[(A4)]For any $\param \in \paramspa_{m}$ and $\param' \in \paramspa_{m'}$, $\mathbb{V}_{y'\sim f(\cdot\mid x')}\left[\frac{q_{m}(y'\mid \param)}{q_{m'}(y'\mid \param')}\right] < \infty\,,\forall m,m'$.
\end{description}
These assumptions are met when $\paramspa_m$ is a bounded set for each and every model in the set, such that
\begin{align*}
R_m := \sup^{}_{\param_{m}\in\paramspa_{m}} \|\param_{m}\|,\quad S_m:=\sup^{}_{y\in\mathcal{Y}} \|s_m(y)\|,\quad K_m:=\exp\{R_m S_m\}
\end{align*}
are finite, which gives $0< \exp\{-R_m S_m\} \leq q_{m}(y\mid \param_{m}) \leq \exp\{R_m S_m\}$ for any $\param_{m}=\param$ and $y\in\mathcal{Y}$, using the Cauchy-Schwartz inequality. 
Hence, we can set $\mathbb{V}_{y'\sim f(\cdot\mid x')}\left[\frac{q_{m}(y'\mid \param)}{q_{m'}(y'\mid \param')}\right] \leq K_m K_{m'}$.

We acknowledge that Assumptions (A1) to (A4) are, in general, quite strong. If $\Theta_m$ is not a bounded set then they will be unrealistic.
However, for the specific case of exponential random graph models, these assumptions can be 
deemed realistic for the following reason. These models are known to feature degenerate regions, i.e. parts of the parameter space where the model generates 
only full or empty graphs. Hence, the model parameters will effectively lie on a compact space and so a Bayesian analysis of those models may be carried out, 
as a matter of course, with priors whose support is included in the non-degenerate region. In this sense, we feel that Assumptions (A1) to (A4) become rather
mild so that the theoretical results are then relevant. 
\begin{lemma}\label{lemma:noisy_exchange_bound_SIS}
Under assumptions (A1)-(A4), for any $x,x'\in\mathcal{X}$, $\hat{\rho}(x,x',y')$ estimated with \eqref{eqn:biased_IS} satisfies 
\begin{align*}
\mathbb{E}_{y'\sim f(\cdot\mid x')}\left|\hat{\rho}(x,x',y')-\rho(x,x')\right|& \leq \gamma^{}_{IS}(x,x')\\
& =
\frac{1}{\sqrt{N}}
\frac{q_{m'}(y\mid\param')}{q_m(y\mid\param)}
\frac{p_{m'}(\param')}{p_{m}(\param)}
\frac{p(\models_{m'})}{p(\models_m)}
\frac{h(\param,m\mid\param',m')}{h(\param',m'\mid\param,m)}\\
& \mathrel{\phantom{=}}
\times
\sqrt{\mathbb{V}_{y'\sim f(\cdot\mid x')}\left[\frac{q_{m}(y'\mid \param)}{q_{m'}(y'\mid \param')}\right]},
\end{align*}
where the upper bound $\gamma^{}_{IS}(x,x')$ refers to the case when the ISE is used.
\end{lemma}
Theorem \ref{theorem:noisy_exchange_bound_SIS} places a bound on the total variation between the Markov chain of a noisy RJMCMC algorithm and a Markov chain with the desired target distribution, when the intractable ratio of normalising constants is estimated by the ISE in \eqref{eqn:biased_IS}. Theorem 3.1 in \cite{alquier} is the special case of Theorem \ref{theorem:noisy_exchange_bound_SIS} when a within-model transition is attempted.
\begin{theorem}\label{theorem:noisy_exchange_bound_SIS}
Under assumptions (A1)-(A3) then (H2) in Corollary \ref{cor:noisy_rjmcmc_bound} is satisfied with
\begin{equation*}
\gamma^{}_{IS}(x,x')
\leq
\frac{c^{}_{p_m}c^{}_{p_{m'}}c^{}_{\models_m}c^{}_{\models_{m'}}c^{}_{\upsilon(x,x')}c^{}_{\upsilon(x',x)}K^2_m K^2_{m'}}{\sqrt{N}}
\end{equation*}
and
\begin{equation*}
\|\delta_{x_0}\mathcal{P}^n - \delta_{x_0}\hat{\mathcal{P}}^n\|
\leq
\frac{\mathcal{D}^{}_{IS}}{\sqrt{N}},
\end{equation*}
where $\mathcal{D}^{}_{IS}=\mathcal{D}^{}_{IS}(c_{p},c_{\models},c_{h},K)$ is explicitly known, see \eqref{eq:UB_IS} of the Appendix.
\end{theorem}
Empirical results in this paper have demonstrated that the use of the TPE leads to an improved RJMCMC algorithm relative to the RJ exchange algorithm. We present the following Lemma that shows how to control the variance of the TPE with $L,S>1$.
\begin{lemma}[Lemma 6 in \cite{boland}]\label{lemma:var_prod_indep_rvs}
The TPE requires $L-$1 sets of simulated data,
\begin{equation*}
y'=
\begin{matrix}
y^{(2)'}  & = &y^{(2)'}_1,&\ldots, &y^{(2)'}_S \sim f_{m'}(\cdot\mid \param'(t^{}_{2}))\\
\vdots&   &\vdots    &\ddots  &\vdots\\
y^{(L)'}  & = &y^{(L)'}_1,&\ldots, &y^{(L)'}_S \sim f_{m'}(\cdot\mid \param'(t^{}_{L})),
\end{matrix}
\end{equation*}
that are used to estimate the ratio of normalising constants with \eqref{eqn:PSE_ratio}. Let $\overline{X}^1_S,\ldots,\overline{X}^{L-1}_S$ be $L-1$ i.i.d one-dimensional sample mean estimators, such that
\begin{equation*}
\overline{X}^j_S=\frac{1}{S}\sum_{b=1}^{S}\frac{q_m(y^{(j+1)'}_b\mid \param'(t^{}_j)))}{q_{m'}(y^{(j+1)'}_b\mid \param'(t^{}_{j+1}))},~~
y^{(j+1)'}_1,\dots,y^{(j+1)'}_S \overset{i.i.d}{\sim} f_{m'}(\cdot\mid \param'(t^{}_{j+1})),\,j\in\{1,\ldots,L-1\}.
\end{equation*}
When assumptions (A1)-(A4) are satisfied, then
\begin{equation*}
\var_{y'}\left[\prod_{j=1}^{L-1}\overline{X}^j_S\right] \leq [K_m K_{m'}]^{L-1}\left\{\left(1+\frac{1}{S}\right)^{L-1} -1\right\}.
\end{equation*}
\end{lemma}
Lemma \ref{lemma:var_prod_indep_rvs} provides guidelines regarding the rate of convergence of $L,S$ to infinity. It shows that $S$ should converge to infinity at a faster rate than $L$, so that $1/S\rightarrow 0$. Of course, increasing the number of steps, $L$, in the transition path allows for points $\param'(t^{}_j),\param'(t^{}_{j+1}), j=1,\ldots L-1$ that are closer together in terms of the $L^2$ norm. 
This, in turn, means that the distribution $f_{m'}(y'\mid\param'(t^{}_{j+1}))$ is only slightly different from $f_{m'}(y'\mid\param'(t^{}_{j}))$ and serves as an excellent importance distribution. In both ERGM examples we show that even smaller values of $L,S$ can lead to good performance of the noisy RJMCMC algorithm. These values can increase according to the available computational resources.

Below we provide a bound on the total variation distance between the Markov chain of a noisy RJMCMC algorithm with the TPE and a Markov chain with the desired target distribution with respect to the factors $L,S$.
\begin{lemma}\label{lemma:noisy_exchange_bound_TPE}
Under assumptions (A1)-(A4), for any $x,x'\in\mathcal{X}$, $\hat{\rho}(x,x',y')$ estimated with \eqref{eqn:PSE_ratio} satisfies 
\begin{align*}
\mathbb{E}_{y'\sim f(\cdot\mid x')}\left|\hat{\rho}(x,x',y')-\rho(x,x')\right|& \leq \gamma^{}_{TP}(x,x')\\
& =
\frac{q_{m'}(y\mid\param')}{q_m(y\mid\param)}
\frac{p_{m'}(\param')}{p_{m}(\param)}
\frac{p(\models_{m'})}{p(\models_m)}
\frac{h(\param,m\mid\param',m')}{h(\param',m'\mid\param,m)}\\
& \mathrel{\phantom{=}}
\times
\sqrt{\mathbb{V}_{y'}\left[\prod_{j=1}^{L-1}\frac{1}{S}\sum_{b=1}^{S}\frac{q_m(y^{(j+1)'}_b\mid \param'(t^{}_j)))}{q_{m'}(y^{(j+1)'}_b\mid \param'(t^{}_{j+1}))}\right]},
\end{align*}
where the upper bound $\gamma^{}_{TP}(x,x')$ refers to the case when the TPE is used.
\end{lemma}
We conclude by an application of Corollary \ref{cor:noisy_rjmcmc_bound} that allows to assess the convergence of the noisy RJMCMC scheme with the TPE for $L,S\gg 1$.
\begin{theorem}\label{theorem:noisy_exchange_bound_TPE}
Under assumptions (A1)-(A3) then (H2) in Corollary \ref{cor:noisy_rjmcmc_bound} is satisfied with
\begin{equation*}
\gamma^{}_{TP}(x,x')\leq 
c^{}_{p_m}c^{}_{p_{m'}}
c^{}_{\models_m}c^{}_{\models_{m'}}
c^{}_{\upsilon(x,x')}c^{}_{\upsilon(x',x)}
[K_m K_{m'}]^{\frac{L+1}{2}}\left\{\left(1+\frac{1}{S}\right)^{L-1} -1\right\}^{1/2}
\end{equation*}
and
\begin{equation*}
\|\delta_{x_0}\mathcal{P}^n - \delta_{x_0}\hat{\mathcal{P}}^n\|
\leq
\mathcal{D}^{}_{TP}
[K_m K_{m'}]^{\frac{L-1}{2}}\left\{\left(1+\frac{1}{S}\right)^{L-1} -1\right\}^{1/2},
\end{equation*}
where $\mathcal{D}^{}_{TP}=\mathcal{D}^{}_{TP}(c^{}_{p},c^{}_{\models},c^{}_{h},K)$ is explicitly known, see \eqref{eq:UB_TP} of the Appendix.
\end{theorem}

\section{Applications to exponential random graph models}\label{section:applications}
We apply the reversible jump methodology to some challenging models for the analysis of network data. All computations in this paper were carried out 
with the statistical environment \pkg{R} \citep{rsoft}.

The two ERGM examples below aim at providing insights regarding the efficiency of the noisy RJMCMC algorithm relative to the RJ exchange algorithm. 
Any RJMCMC algorithm allows for Bayesian multi-model inference as well as estimation of posterior model probabilities and so we are also interested in the performance of the RJMCMC algorithm for each model separately.
We assess this performance in terms of:
\begin{itemize}
	\item The estimated within-model acceptance rate and effective sample size ($\text{ESS}^{}_{W}$) of the sampler for the model with the highest estimated posterior model probability. For an MCMC run of length $G$ with lag $\ell$ auto-correlation $\rho_{\ell}$ the effective sample size is defined as $\text{ESS}^{}_{W}=G/(1 + 2\sum\limits_{\ell=1}^{\infinity}\rho_{\ell})$ \citep{liu}. 
	This $\text{ESS}^{}_{W}$ (the larger the better) gives an estimate of the equivalent number of independent iterations that the chain represents. In our implementation of the RJMCMC algorithms, $\text{ESS}^{}_{W}$ is estimated separately for each parameter and so we report the smallest estimate among these.
	\item The estimated Bayes factor, between-model acceptance rate and total computation time.
  \item $\text{ESS}^{}_{B}$, the estimated effective sample size of the Markov chain that targets the discrete stationary distribution with probabilities $(\pi(\models_1\mid y),\pi(\models_2\mid y),\pi(\models_3\mid y),\ldots)$. We implemented the computational approach proposed in \cite{heck} to estimate $\text{ESS}^{}_{B}$ and define the efficiency (EFF) of each RJMCMC algorithm as $\text{ESS}^{}_{B}$ per second.
\end{itemize}
Throughout the analysis we chose a Multivariate Gaussian prior distribution for the model-specific parameters, $\mathcal{MVN}\left(0_{d_m},10\mathcal{I}_{d_m}\right)$, where $0_{d_m}$ is the null vector and $I_{d_m}$ is the identity matrix of size equal to the number of model dimensions $d_m$. 
In further simulation experiments not shown here, changing the prior distribution to $\mathcal{MVN}\left(0_{d_m},100\mathcal{I}_{d_m}\right)$ did not lead to very different results.
The parameter spaces $\Theta_m, m\in\{1,2,3,\ldots\}$ were defined on a bounded set for both examples, so that Assumptions (A1) to (A4) hold.
We assumed that the models are equally probable a-priori.
Sampling from the likelihood at each RJMCMC iteration was performed with an auxiliary Markov chain of length 3,000 as a proxy for an exact sampler. 
Each main Markov chain was run for 500,000 iterations discarding the first 50,000 as burn-in. This allowed for adequate exploration of the posterior distribution of each model.

The Auto-RJ sampler performed a pilot MCMC run for each model using the population MCMC approach of \cite{caimo}, with $2000\times d_m$ main iterations (discarding the first $500\times d_m$ iterations as burn-in). The Metropolis tuning parameters involved in the RW RJMCMC algorithms were appropriately selected to give the desired within-model acceptance rates.

Finally, the TPE was obtained using a linear path (equal spacing) in $t\in[0,1]$ that consisted of $L$ steps, including the starting and finishing point. 
At each path point the auxiliary step to draw $y'$ was followed by an extra number of iterations thinned by a factor of 50, yielding $S$ graphs. In Section \ref{section:karate} the TPE was implemented using serial computation on a single core for illustration purposes. Of course, this adds to the CPU burden per iteration of the noisy RJMCMC algorithm if a more refined transition path is considered. 
Assuming that parallel computations occur no additional cost, the $L-1$ forward-simulations required in \eqref{eqn:PSE_ratio} can be performed in parallel, taking advantage of the inherently parallel nature of the TPE while yielding a useful noisy RJMCMC algorithm. We implemented a parallel computation of the TPE in Section \ref{section:lazega} to emphasise the improvements offered by a parallel implementation of the TPE relative to the RJ exchange algorithm.

\subsection{Exponential random graph models} \label{section:ergm}
Networks are relational data represented as mathematical graphs of nodes and edges. A $n\times n$ random adjacency matrix \textit{Y} on \textit{n} nodes and a set of edges (relationships) describes the connectivity pattern of a graph, considering the set of all possible graphs on \textit{n} nodes (actors), $\mathcal{Y}$. A realisation of $\textit{Y}$ is denoted with $y$ and the presence or absence of an edge (directed or undirected) between the pair of nodes $(i,j)$ is coded as
\begin{equation*}
  y_{ij}=\begin{cases}
	        1, &\textrm{if }(i,j) \textrm{ are connected,}\\
          0, &\textrm{otherwise.}\\
          \end{cases}
\end{equation*}
An edge connecting a node to itself is not permitted so $y_{ii} = 0$.

Exponential random graph models play an important role in network analysis since they can represent transitivity and other structural features in network data that define complicated dependence patterns not easily modeled by more basic probability models \citep{wasserman}. 
The reader is also referred to \cite{robins2} for a detailed review and the references therein for more details. ERGMs belong to the exponential family of distributions with natural parameter $\param$ and sufficient statistics $s(y)$ and are Gibbs random fields that are defined on the edge space of networks. 

The distribution of $\textit{Y}$ assuming model $\models_m$ is formulated as
\begin{align} \label{eqn:ergmprob}
f_m(y\mid\theta_m)=\frac{q_m(y\mid\param_m)}{z_m({\param_m)}}=\frac{\exp\left\{\param^\top_m s_m(y)\right\}} {\sum_{y\in \mathcal{Y}}^{} \exp\left\{\param^\top s_m(y)\right\}},\qquad \param_m^\top s_m(y)=\sum_{j=1}^{d_m}\param_{mj} s_{mj}(y),
\end{align}
where $q_m(y\mid\param_m)$ is the un-normalised likelihood, $s_m(y)\in \mathcal{S}_m\subseteq\mathbb{R}^{d_m}_{+}$ is a known vector of overlapping sub-graph configurations/ sufficient statistics (e.g. the number of edges, degree statistics, triangles, etc.) and $\param_m\in\paramspa_m\subseteq\mathbb{R}^{d_m}$ is the vector of model parameters. 
More examples can be found at \cite{snijders_recent_2007} and \cite{hunter}. 
Additionally, ERGMs allow for inclusion of covariate information $\textit{X}$, eg. the number of edges on the graph within the same attribute category as a measure of the homophily effect of that attribute. The respective sufficient statistics are denoted by $s_m(y,x)$.

The evaluation of $z_m(\param_m)$ is feasible for only trivially small graphs as this sum involves $2^{\binom{n}{2}}$ terms for undirected edge graphs and $2^{n(n-1)}$ terms for directed edge graphs. Recent works on the inference of ERGMs with the Bayesian approach have been proposed by \cite{koskinen}, \cite{caimo}, \cite{atchade}, \cite{caimo4}, \cite{thiemichen} and \cite{bouranismisp}.
\begin{remark}
The estimators \eqref{eqn:biased_IS} and \eqref{eqn:PSE_ratio} require perfect sampling from the likelihood to generate the auxiliary data. 
This is feasible for GRFs such as the Ising and Potts models \citep{propp,huber1}, but this is not the case for ERG models. 
\cite{caimo} considered the pragmatic alternative of replacing exact simulation with Gibbs updates, where a long auxiliary Markov chain implemented by the tie-no-tie (TNT) sampler is used to return $N=1$ draw that is approximately distributed under the true likelihood, in place of exact simulation \citep{ergm}. 
It is possible to carry out inference for graphs of larger size (eg. 1000 nodes), but at the cost of an increased computational time. 
For convergence results of the Markov chain that uses approximate draws from the likelihood, see \cite{Everitt} and \cite{atchade}.
\end{remark}
Let $aux$ be the number of auxiliary iterations used in the TNT sampler and let $k$ be the lag between consecutive draws from the target distribution of the sampler.
The computational cost per iteration of the noisy RJMCMC algorithm for the estimation of the ratio $z_m(\param)/z_{m'}(\param')$ is $\mathcal{O}(aux+kN)$ when the ISE is used and $\mathcal{O}((L-1)(aux+kS))$  when the TPE is used. 
For the special case of the RJ exchange algorithm, the computational cost per iteration is $\mathcal{O}(aux+1)$. 
For models like GRFs, these estimators depend on the simulated data only through the collection of sufficient statistics $s_{m'}(y)$ which need to be stored in the computer memory.

\subsection{Zachary's karate club network}\label{section:karate}
\begin{figure}[H]
\centering
\includegraphics[clip,trim={0 0 0 2cm},width=8cm,height=8cm]{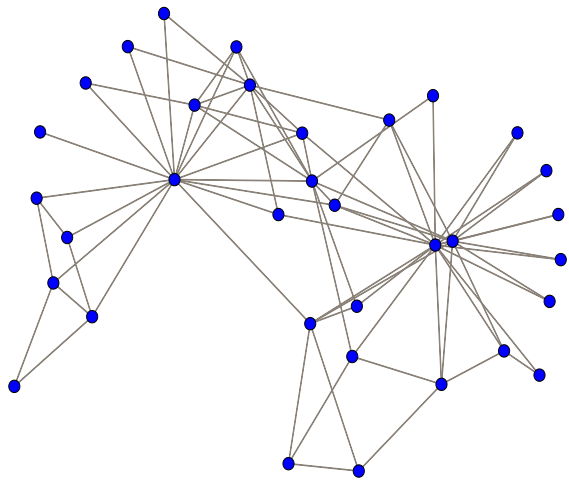}
\vspace{-2.6em} 
\caption{Zachary's Karate Club graph.}
\label{fig:karate_club_graph}
\end{figure}

Zachary's Karate Club (Figure \ref{fig:karate_club_graph}) is a social network of friendships between 34 members of a karate club at a US university in 1970. For this network we are interested in the effect of triad closure, therefore it is natural to consider $k$-star statistics for $k\geq2$ and triangle counts, which express the level of transitivity.

For a fixed network density however, $k$-stars become more prevalent as heterogeneity increases at the cost of often inducing degeneracy. 
The \textit{degeneracy problem} causes the generated graphs to be either very sparse or very dense and only in rare cases does a generated graph have a density close to that of the data network. 
By considering some commonly used network statistics like the heterogeneity of degree statistics and high-order transitivity statistics \citep{snijders_recent_2007,hunter2}, we try to remedy the degeneracy of $k$-stars and triangle counts. Two competing models are assessed:
\begin{center}
\addtolength{\tabcolsep}{-3pt}    
\begin{tabular}{ll}
$\models_1$: & $q_1(y\mid \param_1)=\exp\big\{\param_{11}s_1(y)+\param_{12}v(y,\phi_v)\big\}$ \\
$\models_2$: & $q_2(y\mid \param_2)=\exp\big\{\param_{21}s_1(y)+\param_{22}v(y,\phi_v)+\param_{23}u(y,\phi_u)\big\}$, \\
\end{tabular}
\addtolength{\tabcolsep}{3pt}
\end{center}
where $s_1(y)=\sum_{i<j}^{}y_{ij}$ is the number of edges. The other model terms are defined below.
\vspace{-1.0em}

\begin{description}
\item[Shared Partnership:]
Let $EP_k(y)$, called the edgewise shared partnership statistic, denote the number of connected pairs with exactly $k$ common neighbors. $EP_k(y)$ is a function of the triangle counts and as such, it is equivalent to modeling the high-order transitivities. 
The distribution of edgewise shared partnership can be modeled as a function of a single parameter by placing decreasing weights on the higher transitivities, leading to the geometrically weighted edgewise shared partnership (GWESP) statistic. GWESP is defined by:
\begin{align*}
v(y,\phi_v) = e^{\phi_v} \sum_{k=1}^{n-2}\left\{ 1-\left(1 - e^{-\phi_v} \right)^{k}\right\}EP_k(y)\,.
\end{align*}
\item[Geometrically Weighted Degree:]
 Let the degree count, $D_k(y)$, denote the number of pairs that have exactly $k$ common neighbors. 
The number of stars is a function of the degrees, therefore $D_k(y)$ is equivalent to modeling the $k$-star statistic. 
The geometrically weighted degree (GWD) statistic enables to model all degree distributions as a function of single parameter by placing decreasing weights on the higher degrees. GWD is defined by:
\begin{align*}
u(y,\phi_u) = e^{\phi_u} \sum_{k=1}^{n-1}\left\{ 1-\left(1 - e^{-\phi_u} \right)^{k}\right\}D_k(y)\,.
\end{align*}
\end{description}
The scale parameters $(\phi_v\,,\phi_u)$ specify the decreasing rates of weights placed on the higher order terms, are treated as constants and are set to $(\phi_{v}\,,\phi_{u}) = (0.2\,,0.8)$, following \cite{caimo2}. 
Tables \ref{tab:karate_example_BF_summary} points towards positive evidence in favor of $\models_1$ over $\models_2$. This shows that the effect captured by the geometrically weighted degree network statistic does not enhance the observed network.

Some general comments can be made about the performance of the two algorithms. In this example all algorithms suffer from low acceptance rates of the trans-dimensional move, since the probability mass is concentrated on one model. 
It would be expected that the proposed jumps would be accepted more frequently if the probability mass was spread over different models.
The Auto-RJ sampler for this example yields a Markov chain with a better asymptotic efficiency than the RW RJMCMC algorithm, however this algorithm will become costly when 
the number of models is large as it can be computationally expensive to tune the proposal distributions for each competing model. 
In terms of asymptotic efficiency the Auto-RJ sampler, in general, may become way worse than the noisy RW RJMCMC sampler simply because of the fact that the independent Gaussian 
proposal may not be a good match to the posterior distribution for some of the competing models.
\begin{table}[H]
\caption{Zachary karate club - Bayes factor (standard deviation), acceptance rate of the trans-dimensional move and CPU time in hours based on thirty independent noisy RJMCMC runs. 
We also report 
the effective sample size of the Markov chain on the posterior model probabilities ($\text{ESS}^{}_{B}$) and the efficiency (EFF) for each algorithm.
}
\centering
\vspace{-0.7em}
\begin{tabular}{@{\extracolsep{-2.3pt}}lrrrrrr@{}}
\toprule
               &  &\multicolumn{3}{c}{\textbf{ISE $\pmb{(N)}$}}        & \multicolumn{2}{c}{\textbf{TPE $\pmb{(L\,,S)}$}}\\
\cline{3-7} \noalign{\vskip 0.2em}
\textbf{Method}&  & $\mathbf{1}$      & $\mathbf{5}$ & $\mathbf{10^2}$ & \textbf{(6,1)}  & \textbf{(11,10)}\\
\hline
RW-RJ  & $BF_{12}$     & 12.81 (0.37) & 10.67 (0.32) & 13.53 (0.33)& 13.01 (0.42) & 13.07 (0.45)  \\
       & \% accepted   & 6.3          & 9.8          & 6.9         & 7.2          & 7.5   \\
			 & $\text{ESS}^{}_{B}$& 69,479 & 78,575 & 85,389 & 89,612 &  93,090\\ 
			 & CPU                & 0.92    & 0.98          & 2.04        & 4.08 & 8.95   \\
			 & EFF                & 18.94  & 22.27  & 10.69  & 5.15   &  2.65 \\
\hline
Auto-RJ& $BF_{12}$     & 13.06 (0.21)  & 15.19 (0.24)  & 15.21 (0.26)   & 13.08 (0.17) & 13.09 (0.15)\\
       & \% accepted   & 5.6           & 6.6           & 8.3            & 9.6          & 10.6 \\
			 & $\text{ESS}^{}_{B}$& 53,760   & 75,251 & 98,852  & 100,101 & 113,297 \\ 
			 & CPU                & 1.76  & 1.83  & 2.22   & 4.49 & 9.88 \\
			 & EFF                & 8.48     & 18.41  & 12.37   & 6.19    & 1.66 \\
\bottomrule
\end{tabular}
\label{tab:karate_example_BF_summary}
\end{table}
Table \ref{tab:karate_example_BF_summary} shows how all of the noisy RJMCMC algorithms displayed better mixing in terms of acceptance rate of the trans-dimensional move when compared to the RJ exchange algorithm. 
This improvement, though, comes with a cost: an ISE based on $N>1$ appears to have an effect on the frequency of visits to each model, hence biasing the Bayes factor estimate. All samplers are faced with a biased Bayes factor estimate when $N=5$ and this bias appears to decrease with larger $N$, when the noisy acceptance ratio mimics the MH acceptance ratio.

The noisy RJMCMC samplers also benefit from improved performance in terms of mixing when the TPE is used for the estimation of the intractable acceptance ratio. 
Table \ref{tab:karate_example_BF_summary} suggests that even a small number of draws from the auxiliary distribution can improve the acceptance rate of the trans-dimensional move, resulting in a 40\% increase on average for the noisy Auto-RJ {sampler $(L=6,\,S=1)$ relative to the Auto-RJ exchange algorithm ($N=1)$}, at the cost of some additional computational expense. 
Increasing the overall number of auxiliary draws further improves the mixing of the noisy RJMCMC algorithms, but then their implementation becomes impractical even for this small model set. This is an indicator that the TPE will have small variance, making it particularly useful for RJMCMC.

Table \ref{tab:karate_example_AR_summary} shows an improvement in the ESS for the most probable model over the RJ exchange algorithm ($N=1$), which translates into a decrease in the auto-correlation of the chain for each model. 
Both the ISE and the TPE improved the mixing within each model, which is in line with the results of \cite{alquier}. 
Taken together with the results of Table \ref{tab:karate_example_BF_summary}, we can observe a trade-off between bias and efficiency, as the noisy RJMCMC algorithm is targeting an approximation of the true posterior \eqref{eq:joint_posterior}.

A natural question that arises is how big should $N$ or $L,S$ be in order to have a trade-off between the bias of the distribution and the computational time. 
From this experiment, it appears that the bias of the Bayes factor estimate is connected to the bias in the target distribution of the noisy RJMCMC chain. Knowledge of the variance of the estimator of the ratio of intractable ratio of normalising constants could give a rough idea about choosing appropriate values for the factors $N$ or $L,S$.

Finally, the results in Table \ref{tab:karate_example_BF_summary} point towards decreased efficiency of the noisy RJMCMC algorithm when a larger number of likelihood draws ($N=100$) is taken. As discussed in \cite{alquier}, the rationale behind the introduction of a noisy (RJ)MCMC algorithm is that the combined statistical and computational efficiency is sufficiently improved to outweigh the effect of any bias that is introduced. In the case of an ISE with $N=100$, however, the increased CPU cost decreases the efficiency of the algorithm. 
A serial implementation of the TPE decreases the efficiency of the noisy RJMCMC algorithm relative to the RJ exchange sampler, as expected. In Section \ref{section:lazega} we show that a parallel implementation of the TPE helps to improve the efficiency.

\begin{table}[H]
\caption{Zachary karate club - Acceptance rate and effective sample size of the Markov chain on the posterior parameter estimates ($\text{ESS}^{}_{W}$) for the most probable model, $\models_1$, based on thirty independent noisy RJMCMC runs.
}
\centering
\vspace{-0.7em}
\begin{tabular}{llrrrrr}
\toprule
&                 &\multicolumn{3}{c}{\textbf{ISE $\pmb{(N)}$}}   & \multicolumn{2}{c}{\textbf{TPE $\pmb{(L\,,S)}$}} \\
\cline{3-7} \noalign{\vskip 0.2em}
\textbf{Method}&  & $\mathbf{1}$ & $\mathbf{5}$ & $\mathbf{10^2}$ & \textbf{(6,1)} & \textbf{(11,10)} \\
\hline
RW-RJ 
& AR  & 22\%  & 21\%  & 22\%  & 21\%   & 22\%\\
& $\text{ESS}^{}_{W}$ & 12,498& 16,094& 20,297& 20,664 & 24,199 \\
\hline
Auto-RJ
& AR  & 24\%   & 28\%   & 38\%   & 40\%   & 46\% \\
& $\text{ESS}^{}_{W}$ & 18,855 & 25,744 & 43,950 & 47,153 & 62,795  \\
\bottomrule
\end{tabular}
\label{tab:karate_example_AR_summary}
\end{table}

\subsection{Collaboration between Lazega's lawyers}\label{section:lazega}
\begin{figure}[H]
\centering
\includegraphics[clip,trim={0 0 0 2cm},width=8cm,height=8cm]{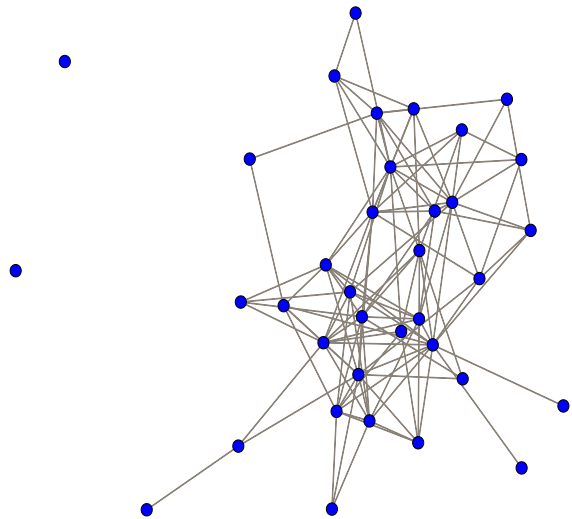}
\vspace{-2.6em} 
\caption{Lazega's network of corporate law partnership.}
\label{fig:lazega_graph}
\end{figure}

The Lazega network dataset (Figure \ref{fig:lazega_graph}) originates from a network study of corporate law partnership that was carried out in a Northeastern US corporate law firm in New England \citep{lazega}. 
The dataset consists of $36$ nodes (partners) and the presence of an edge between two nodes indicates a collaboration between the two partners. Information about nodal attributes is also available; here we are interested in the attribute variables of gender (1=male; 2=female) and practice (1=litigation; 2=corporate).

We compare the following three models as in \cite{caimo2}: 
\begin{center}
\addtolength{\tabcolsep}{-3pt}    
\begin{tabular}{ll}
$\models_1$: & $q_1(y\mid \param_1)=\exp\big\{\param_{11}s_1(y)+\param_{12}v(y,\phi_v)+\param_{13}s_{1}(y,x)+\param_{14}s_{2}(y,x)+\param_{15}s_{3}(y,x)\big\}$ \\
$\models_2$: & $q_2(y\mid \param_2)=\exp\big\{\param_{21}s_1(y)+\param_{22}v(y,\phi_v)+\param_{23}s_{1}(y,x)+\param_{24}s_{2}(y,x)\big\}$ \\
$\models_3$: & $q_3(y\mid \param_3)=\exp\big\{\param_{31}s_1(y)+\param_{32}v(y,\phi_v)\big\}$ \\
\end{tabular}
\addtolength{\tabcolsep}{3pt}
\end{center}
where $s_1(y)$ and $v(y,\phi_v)$ are the same terms as in Section \ref{section:karate} and $\phi_v=\log(2)$. The covariate statistics are defined as
\begin{equation*}
s(y,x) = \sum_{i\neq j} y_{ij}\times
\begin{dcases}
x_{i} + x_{j}                   & \text{: "main effect"}, \\
\mathbbm{1}_{\{x_{i} = x_{j}\}} & \text{: "homophily effect"}. 
\end{dcases}
\end{equation*}
The covariate statistic $s_{1}(y,x)$ represents the homophily effect of practice, $s_{2}(y,x)$ is the homophily effect of gender and $s_{3}(y,x)$ is the main effect of practice.
\vspace{-1.0em}
All algorithms point towards $\models_3$ having the highest estimated posterior model probability. For illustration purposes, below we report the estimated Bayes factor for the comparison between $\models_3$ against $\models_2$. The results in Table \ref{tab:lazega_example_BF_summary} show that there is positive evidence in favor of $\models_3$ against $\models_2$.
\begin{table}[H]
\caption{Lazega's lawyers - Bayes factor, acceptance rate of the trans-dimensional move, effective sample size of the Markov chain on the posterior model probabilities ($\text{ESS}_{pm}$) and efficiency (EFF) for each algorithm, based on a long RJMCMC run.
}
\centering
\vspace{-0.7em}
\begin{tabular}{@{\extracolsep{-2.3pt}}lrrrrrr@{}}
\toprule
               &  &\multicolumn{3}{c}{\textbf{ISE $\pmb{(N)}$}}        & \multicolumn{2}{c}{\textbf{TPE $\pmb{(L\,,S)}$}}\\
\cline{3-7} \noalign{\vskip 0.2em}
\textbf{Method}&  & $\mathbf{1}$      & $\mathbf{5}$ & $\mathbf{10^2}$ & \textbf{(6,1)}  & \textbf{(11,10)}\\
\hline
RW-RJ  & $BF_{32}$        & 13.12  & 23.12 & 13.01 & 13.35 & 13.76  \\
       & \% accepted      & 1.0    & 1.2   & 1.3   & 2.0   & 3.0    \\
			 & $\text{ESS}_{pm}$& 1,007  & 4,131 & 5,382 & 5,104 & 6,422  \\
			 & EFF              & 0.23   & 0.87  & 1.25  & 1.16  & 1.46   \\
\hline
Auto-RJ& $BF_{32}$        & 13.46  & 46.31  & 13.72  & 14.02 & 13.89  \\
       & \% accepted      & 1.0    & 1.0    & 2.0    & 3.0   & 3.0    \\
			 & $\text{ESS}_{pm}$& 11,211 & 59,200 & 58,863 & 41,432& 54,146 \\ 
			 & EFF              & 1.26   & 6.55   & 4.65   & 4.66  & 6.05   \\ 
\bottomrule
\end{tabular}
\label{tab:lazega_example_BF_summary}
\end{table}
The higher-dimensionality of the competing models allows us to investigate the performance of noisy RJMCMC in a situation that is more likely to be faced in practice. 
Table \ref{tab:lazega_example_BF_summary} shows that a small number of draws on the ISE ($N=5$) leads to a bias in the Bayes factor estimate that is pronounced in this example for both the random walk and the independence samplers.
At the same time, the mixing of the RJ Markov chain improves, but many more draws ($N>100$) will be needed to obtain a better approximation to the target posterior distribution and consequently reduce the bias in the estimated posterior model probabilities and the Bayes factor estimate.

On the contrary, the TPE proves itself useful even with a small number in the factors $L,S$, eg. $(L=6,\,S=1)$.
The increased efficiency of the noisy RJMCMC algorithms with the parallel implementation of the TPE depicted in Table \ref{tab:lazega_example_BF_summary} shows more emphatically that the new approaches offer large improvements over running the RJ exchange algorithm, while simultaneously improving the statistical ($\text{ESS}_{pm}$) and computational (EFF) efficiency of the sampler.

The same comments made for Table \ref{tab:karate_example_AR_summary} can be also made for the results in Table \ref{tab:lazega_example_AR_summary}: the noisy RJMCMC algorithms had less auto-correlation within each model compared to the RJ exchange algorithm, similarly to Zachary's Karate Club network example.
 
In both ERGM examples, we observed that the noisy independence sampler does better than the noisy RW RJMCMC algorithm. This is a result of having a posterior distribution 
for each model that is close to Gaussian and so the automated design of the independence proposal is efficient. However, this may not generally be the case and in this instance we expect that the noisy RW RJMCMC algorithm may give better results.
\begin{table}[H]
\caption{Lazega's lawyers - Acceptance rate and effective sample size of the Markov chain on the posterior parameter estimates ($\text{ESS}^{}_{W}$) for the most probable model, $\models_{3}$, based on a long RJMCMC run.
}
\centering
\vspace{-0.7em}
\begin{tabular}{llrrrrr}
\toprule
&                 &\multicolumn{3}{c}{\textbf{ISE $\pmb{(N)}$}}   & \multicolumn{2}{c}{\textbf{TPE $\pmb{(L\,,S)}$}} \\
\cline{3-7} \noalign{\vskip 0.2em}
\textbf{Method}&  & $\mathbf{1}$ & $\mathbf{5}$ & $\mathbf{10^2}$ & \textbf{(6,1)} & \textbf{(11,10)} \\
\hline
RW-RJ 
& AR                  & 22\%  & 19\%  & 25\%  & 24\%  & 25\% \\
& $\text{ESS}^{}_{W}$ & 4,441 & 4,915 & 5,881 & 9,117 & 10,063\\
\hline
Auto-RJ
& AR                  & 17\%  & 18\%   & 19\%   & 25\%   & 28\% \\
& $\text{ESS}^{}_{W}$ & 8,948 & 10,763 & 15,318 & 20,526 & 25,032\\
\bottomrule
\end{tabular}
\label{tab:lazega_example_AR_summary}
\end{table}

\section{Discussion}\label{section:discussion}
The present paper contributes to the growing literature of approximate MCMC methods for Bayesian analysis of doubly-intractable distributions \citep{alquier,johansen,boland,Everitt2} by introducing a variant of RJMCMC for Bayesian model comparison of Gibbs random fields. 
The resulting algorithm generalises noisy MCMC to trans-dimensional settings, where the transition kernel of the exact RJMCMC algorithm is approximated.
Drawing from the study of the stability of Markov chains \citep{mitrophanov,alquier}, we have given bounds on the total variation between the Markov chain of a 
noisy RJMCMC algorithm and a Markov chain with the desired target distribution in the case where the chain is uniformly ergodic. We acknowledge, though, that this 
is a strong assumption that may not be met in practice.

We have illustrated that the noisy RJMCMC algorithm can suffer from considerable bias (reflected by the Bayes factor estimate) when the variance of the estimator of the ratio of intractable likelihood normalising constants is large. 
In particular, the unbiased estimator proposed in \cite{alquier} is not useful for RJMCMC.
Aiming to overcome these inefficiencies and to decrease the variance of this estimator, we used a variation based on a telescopic product of unbiased importance sampling estimators. 
The telescopic product estimator (TPE) helped reduce the bias of the noisy RJMCMC algorithm, which is then of the same order of magnitude as that of the RJ exchange algorithm, while being more asymptotically efficient than the RJ exchange algorithm. 

Despite the fact that this framework also applies to more general situations apart from Gibbs random field models, there are limitations to the implementation of the noisy RJMCMC algorithm. 
A crucial aspect of the computational performance of the algorithm is its requirement to perform forward simulations from the likelihood at each iteration.
The computational cost is expected to increase when models on large lattices or networks with thousands of nodes are analysed.
The estimators that we investigated in this paper rely on importance sampling and might not be practical in high-dimensional parameter spaces, a point which has been raised by \cite{Everitt} and \cite{Everitt2}. 
In real-life GRF applications, the parameter spaces do not usually exceed 10-15 dimensions.

\section*{Acknowledgements}
\noindent The authors thank the co-editor, the associate editor and the anonymous referees for their constructive comments
that helped improve the article. 
The Insight Centre for Data Analytics is supported by Science Foundation Ireland under Grant Number SFI/12/RC/2289. Nial Friel’s research was also supported by a Science Foundation Ireland grant: 12/IP/1424.

\bibliographystyle{apalike}
\bibliography{CSDA_D1800027_rev2}

\appendix

\section{Proofs}
\begin{proof}[Proof of Proposition \ref{prop:var_bound}]
By definition of $\dot{\param}$, it holds that $z_{m}(\param)=z_{m'}(\dot{\param})$. Then we can write
\begin{align*}
\var_{y'_1, y'_2,\ldots,y'_{N}\sim f_{m'}(\cdot\mid \param')}\left[\widehat{\frac{z_{m'}(\dot{\param})}{z_{m'}(\param')}}\right]
&=
\esp_{\param'}\left[\exp\{2(\dot{\param}-\param')^{\top} s_{m'}(y)\}\right]-\left[\frac{z_{m'}(\dot{\param})}{z_{m'}(\param')}\right]^2 \\
&=
\esp_{\param'}\left[\exp\{2\zeta^{\top} s_{m'}(y)\}\right]-\left[\frac{z_{m'}(\dot{\param})}{z_{m'}(\param')}\right]^2,
\end{align*}
where $\zeta=\dot{\param}-\param'$. We introduce the notation $\psi=\mathcal{O}(\|\zeta\|_2)$, where $\|\cdot\|_2$ is the $L^2$ norm. A Taylor expansion of the variance term around $\param'$ yields
\begin{align*}
\var_{y'_1, y'_2,\ldots,y'_{N}\sim f_{m'}(\cdot\mid \param')}\left[\widehat{\frac{z_{m'}(\dot{\param})}{z_{m'}(\param')}}\right]
&
=1+2\zeta^{\top}\esp_{\param'}\left[s_{m'}(y)\right]+\esp_{\param'}\left[\psi(\zeta,y)\right] \\
& \mathrel{\phantom{=}}
-\frac{1}{z^2_{m'}(\param')}\left[z^2_{m'}(\param')+2z_{m'}(\param')\zeta^{\top}\nabla_{\param'} z_{m'}(\param')+\mathcal{O}(\|\zeta\|^2)\right].
\end{align*}
We note that for models that belong to the exponential family it holds that $\nabla_{\param} z(\param)/z(\param)=\esp_{\param}\left[s(y)\right]~\\\forall\theta\in\paramspa$. Additionally, $\esp_{\param'}\left[\psi(\zeta,y)\right]=\mathcal{O}(\|\zeta\|_2)$, which gives
\begin{equation*}
\var_{y'_1, y'_2,\ldots,y'_{N}\sim f_{m'}(\cdot\mid \param')}\left[\widehat{\frac{z_{m'}(\dot{\param})}{z_{m'}(\param')}}\right]
=\mathcal{O}(\|\zeta\|_2).
\end{equation*}
This concludes the proof.
\end{proof}

\begin{proof}[Proof of Corollary \ref{cor:noisy_rjmcmc_bound}]
Let $x=(m, \param)$ and $x'=(m', \param')$. We assume that there exists a symmetric measure $\mu$ on $\mathcal{X}\times\mathcal{X}$ that dominates
$\pi({\rm d}x)\upsilon(x,{\rm d}x')$. Then $\pi({\rm d}x)\upsilon(x,{\rm d}x')$ has density $\nu(x,x')$ (the Radon-Nikodym derivative) with respect to $\mu$.
For all Borel sets $\mathcal{B,B'}\subset\mathcal{X}$ such that $\mathcal{B'}={\rm d}x'=\{X\in\mathcal{X};\,\|X-x'\|<\epsilon\}$, the following holds for $\mathcal{P}(x,{\rm d}x')$ and $\hat{\mathcal{P}}(x,{\rm d}x')$ in order to apply Theorem \ref{theorem1}:
\begin{equation*}
\mathcal{P}(x,{\rm d}x') =
\upsilon(x,{\rm d}x') \min\left(1,\rho(x,x')\right) + 
\delta_x({\rm d}x') 
\int {\rm d}t \upsilon(x,t)\left[1-\min\left(1,\rho(x,t)\right)\right]
\end{equation*}
and
\begin{align*}
\hat{\mathcal{\mathcal{P}}}(x,{\rm d}x')&=
\int {\rm d}y' f(y'\mid x') \Bigl[ \upsilon(x,x') \min\left(1,\hat{\rho}(x,x',y')\right)\Bigr]
\\
& \mathrel{\phantom{=}}
+
\delta_x({\rm d}x') 
\iint {\rm d}t {\rm d}y'\upsilon(x,t)f(y'\mid t)\left[1-\min\left(1,\hat{\rho}(x,t,y')\right)\right].
\end{align*}
We write
\begin{align*}
(\mathcal{P}-\hat{\mathcal{\mathcal{P}}})(x,{\rm d}x')&=  
\int {\rm d}y' f(y'\mid x')\upsilon(x,{\rm d}x')\Bigl[ \min\left(1,\rho(x,x')\right) - \min\left(1,\hat{\rho}(x,x',y')\right)\Bigr]
\\
& \mathrel{\phantom{=}}
+
\delta_x({\rm d}x')  
\iint {\rm d}t {\rm d}y' \upsilon(x,t) f(y'\mid t)\Bigl[\min\left(1,\hat{\rho}(x,t,y')\right) - \min\left(1,\rho(x,t)\right)\Bigr],
\end{align*}
which gives
\begin{align*}
\|\mathcal{P}-\hat{\mathcal{P}}\|
&= 
\sup_{\substack{x\in\mathcal{X} \\ \mu=\delta_x}}\|\mu(\mathcal{P}-\hat{\mathcal{P}})\|
\\
&= 
\sup_{\substack{x\in\mathcal{X}}}\|\delta_x\mathcal{P}-\delta_x\hat{\mathcal{P}}\|
\\
&=
\frac{1}{2}\sup_{x\in\mathcal{X}} \int  |\mathcal{P}-\hat{\mathcal{\mathcal{P}}}|(x,{\rm d}x')
\\
&= \frac{1}{2}\sup_{x\in\mathcal{X}}\Biggl\{ \Biggl| \iint {\rm d}t {\rm d}y' f(y'\mid t)\upsilon(x,t)\Bigl[\min\left(1,\hat{\rho}(x,t,y')\right) - \min\left(1,\rho(x,t)\right)\Bigr]\Biggr|
\\
& \mathrel{\phantom{=}}
+ \Biggl|\iint{\rm d}y'{\rm d}x' f(y'\mid x')\left[\upsilon(x,x')\min\left(1,\rho(x,x')\right) - \upsilon(x,x') \min\left(1,\hat{\rho}(x,x',y')\right) \right]\Biggr|\Biggr\}
\\
&= 
\sup_{x\in\mathcal{X}} \Biggl\{\Biggl|\iint{\rm d}t {\rm d}y' f(y'\mid t)\upsilon(x,t)\Bigl[\min\left(1,\hat{\rho}(x,t,y')\right) -\min\left(1,\rho(x,t)\right)\Bigr]\Biggr|\Biggr\}
\\
&\leq \sup_{x\in\mathcal{X}}\iint{\rm d}y'{\rm d}x' f(y'\mid x')\upsilon(x,x')\Bigl| \min\left(1,\rho(x,x')\right)- \min\left(1,\hat{\rho}(x,x',y')\right)\Bigr|
\\
&= \sup_{x\in\mathcal{X}}\int{\rm d}x'\upsilon(x,x')\int{\rm d}y' f(y'\mid x')\Bigl| \min(1,\rho(x,x')) - \min(1,\hat{\rho}(x,x',y'))\Bigr|
\\
&\leq \sup_{x\in\mathcal{X}}\int{\rm d}x'\upsilon(x,x')\gamma(x,x').
\end{align*}
\end{proof}

\begin{proof}[Proof of Lemma \ref{lemma:noisy_exchange_bound_SIS}]
Let $x=(m, \param)$ and $x'=(m', \param')$. We check that
\begin{align*}
\mathbb{E}_{y'\sim f(\cdot\mid x')}\left|\hat{\rho}(x,x',y')-\rho(x,x')\right|
&\leq 
\int f(y'\mid x')\left|\hat{\rho}(x,x',y')-\rho(x,x')\right|{\rm d}y' \\
&=
\frac{q_{m'}(y\mid\param')}{q_m(y\mid\param)}
\frac{p_{m'}(\param')}{p_{m}(\param)}
\frac{p(\models_{m'})}{p(\models_m)}
\frac{h(\param,m\mid\param',m')}{h(\param',m'\mid\param,m)} \\
&\mathrel{\phantom{=}} \times
\mathbb{E}_{y_1',\dots,y_N'\sim f(\cdot\mid x')}\left|\frac{1}{N}\sum_{i=1}^{N} \frac{q_{m}(y_i'\mid \param)}{q_{m'}(y_i'\mid \param')} -\frac{z_{m}(\param)}{z_{m'}(\param')}\right| \\
&\leq \frac{1}{\sqrt{N}}
\frac{q_{m'}(y\mid\param')}{q_m(y\mid\param)}
\frac{p_{m'}(\param')}{p_{m}(\param)}
\frac{p(\models_{m'})}{p(\models_m)}
\frac{h(\param,m\mid\param',m')}{h(\param',m'\mid\param,m)} \\
& \mathrel{\phantom{=}}
\times
\sqrt{ \mathbb{V}_{y_1'\sim f(\cdot\mid x')}\left[\frac{q_{m}(y_1'\mid \param)}{q_{m'}(y_1'\mid \param')}\right] }  \\
&=\gamma^{}_{IS}(x,x').
\end{align*}
Then
\begin{align}\label{eq:gamma_SIS_bound}
\gamma^{}_{IS}(x,x')
&\leq
\frac{1}{\sqrt{N}}
c^{}_{p_m}
c^{}_{p_{m'}}
c^{}_{\models_m}
c^{}_{\models_{m'}}
c^{}_{\upsilon(x,x')}
c^{}_{\upsilon(x',x)}
\frac{q_{m'}(y\mid\param')}{q_m(y\mid\param)}
\sqrt{ 
\mathbb{E}_{y'\sim f(\cdot\mid x')}
\left[\left(\frac{q_{m}(y'\mid \param)}{q_{m'}(y'\mid \param')}\right)^2\right] 
} 
\nonumber\\
&\leq
\frac{1}{\sqrt{N}}
c^{}_{p_m}
c^{}_{p_{m'}}
c^{}_{\models_m}
c^{}_{\models_{m'}}
c^{}_{\upsilon(x,x')}
c^{}_{\upsilon(x',x)}
K^{2}_m K^{2}_{m'},
\end{align}
which concludes the proof. 
\end{proof}

\begin{proof}[Proof of Theorem \ref{theorem:noisy_exchange_bound_SIS} (see also Theorem 3.1 of \cite{alquier})]
Under the assumptions of \\ Theorem \ref{theorem:noisy_exchange_bound_SIS}, note that \eqref{eq:rjmcmc_ratio} leads to
\begin{align}\label{proofstep1}
\rho(x,x')
&=
\frac{f_{m'}(y\mid\param')}{f_m(y\mid\param)}
\frac{p_{m'}(\param')}{p_{m}(\param)}
\frac{p(\models_{m'})}{p(\models_m)}
\frac{\omega(m',m)}{\omega(m,m')}
\frac{T_{m',m}(\param', \param)}{T_{m,m'}(\param, \param')} \nonumber\\
&\geq 
[c^{}_{p_m}c^{}_{p_{m'}}
c^{}_{\models_m}c^{}_{\models_{m'}}
c^{}_{\upsilon(x,x')}
c^{}_{\upsilon(x',x)}
K^2_m K^2_{m'}
]^{-1}.
\end{align}
Let us consider any measurable subset $\mathcal{B'}$ of $\mathcal{X}$ and $x'\in\mathcal{X}$. We have
\begin{align*}
\mathcal{P}(x,\mathcal{B'}) 
& = 
\int_{\mathcal{B'}} \delta_x({\rm d}x')\int {\rm d}t \upsilon(x,t)\left[1-\min\left(1,\rho(x,t))\right)\right]
+ \int_{\mathcal{B'}} {\rm d}x' \upsilon(x,x') \min\left(1,\rho(x,x')\right) \\
& \geq 
\int_{\mathcal{B'}} {\rm d}x' \upsilon(x,x') \min\left(1,\rho(x,x')\right)
\\
& \geq 
[c_{p_m}c_{p_{m'}}
c_{\models_m}c_{\models_{m'}}
c_{\upsilon(x,x')}c_{\upsilon(x',x)}
K^2_m K^2_{m'}]^{-1}
\int_{\mathcal{B'}}{\rm d}x'\upsilon(x,x')\text{ thanks to~\eqref{proofstep1}}\\
& \geq 
[c_{p_m}c_{p_{m'}}
c_{\models_m}c_{\models_{m'}}
c^2_{\upsilon(x,x')}c_{\upsilon(x',x)}
K^2_m K^2_{m'}]^{-1} 
\int_{\mathcal{B'}} {\rm d}x'.
\end{align*}
This proves that $\mathcal{B'}$ is a small set for the Lebesgue measure (multiplied by a constant) on $\mathcal{X}$. 
Following the proof of Theorem 3.1 of \cite{alquier} and the reference therein, this proves that:
\begin{equation*}
\sup_{x_0\in\mathcal{X}}\|\delta_{x_0}\mathcal{P}^n-\pi(\cdot\mid y)\|\leq Q\xi^n,
\end{equation*}
where
\begin{equation*}
Q = 2 \text{ and }
\xi = 1 - [c^{}_{p_m}c^{}_{p_{m'}}c^{}_{\models_m}c^{}_{\models_{m'}}c^2_{\upsilon(x,x')}c^{}_{\upsilon(x',x)}K^2_m K^2_{m'}]^{-1}.
\end{equation*}
By definition, $\mathcal{K},c^{}_{p},c^{}_{\upsilon}>1$ and so $\xi\in(0,1)$. This satisfies condition {\bf (H1)} in Corollary \ref{cor:noisy_rjmcmc_bound}. 
Moreover, \eqref{eq:gamma_SIS_bound} satisfies condition {\bf (H2)} in Corollary \ref{cor:noisy_rjmcmc_bound}.
We can apply this Corollary to give
\begin{equation*}
\sup_{x_0\in\mathcal{X}}\|\delta_{x_0}\mathcal{P}^n - \delta_{x_0}\hat{\mathcal{P}}^n\|
\leq 
\frac{\mathcal{D}^{}_{IS}}{\sqrt{N}}
\end{equation*}
with
\begin{equation}\label{eq:UB_IS}
\mathcal{D}^{}_{IS}=c^{}_{p_m}c^{}_{p_{m'}}
c^{}_{\models_m}
c^{}_{\models_{m'}}
c^{}_{\upsilon(x,x')}
c^{}_{\upsilon(x',x)}
K^2_m K^2_{m'}\left(\phi + \frac{Q\xi^{\phi}}{1-\xi}\right),
\end{equation}
where $\phi=\left\lceil \frac{\log(1/Q)}{\log(\xi)} \right\rceil$.
\end{proof}

\begin{proof}[Proof of Lemma \ref{lemma:var_prod_indep_rvs}]
The proof is for trans-dimensional moves and follows from \cite{boland}. We will use the fact that for any one-dimensional random variable $X$,
\begin{equation}
\exists\,M\in\rset\quad\text{s.t.}\quad X\leq M\quad\Rightarrow\quad\var[X]\leq \esp[X^2]\leq M^2.
\end{equation}
By definition of \eqref{eqn:PSE_ratio}, $\var\left[\prod_{j=1}^{L-1}\overline{X}^j_S\right]$ is a collection of $2^{r} - 1$ products of $r=L-1$ positive factors. 
Each factor is either a squared expectation, $\esp^2[\overline{X}^j_S]$, or a variance, $\var[\overline{X}^j_S]$, so that one of the $2^r - 1$ products that contains $k>0$ variances and $r-k$ squared expectations is
\begin{equation*}
p_k:=\prod_{t=1}^{k}\var\left[\overline{X}^t_S\right]\times\prod_{q=k+1}^{r}\esp^2\left[\overline{X}^q_S\right],
\end{equation*}
which can be re-expressed as
\begin{equation*}
p_k:=\frac{1}{S^k}\prod_{t=1}^{k}\var\left[X^t\right]\times\prod_{q=k+1}^{r}\esp^2\left[X^q\right].
\end{equation*}
A uniform bound in $k$ can be placed on $p_k$, such that
\begin{equation}\label{eq:unif_bound}
p_k \leq \frac{1}{S^k}\prod_{t=1}^{k}\esp\left[(X^t)^2\right]\times\prod_{q=k+1}^{r}\esp^2\left[X^q\right]\leq\frac{(K_m K_{m'})^{r}}{S^k}.
\end{equation}
There are $\binom{r}{k}$ terms that have $k$ variances and $r-k$ squared expectations. Therefore, their sum $\overline{p}_k$ can be bounded by the uniform bound in \eqref{eq:unif_bound} so that
\begin{equation*}
\overline{p}_k \leq \frac{(K_m K_{m'})^{r}}{S^k}.
\end{equation*}
Rearrangement of the $2^r-1$ products and aggregation of the products with the same number of factors, $k$, yields
\begin{equation*}
\var\left[\prod_{j=1}^{L-1}\overline{X}^j_S\right]=
\sum_{k=1}^{L-1}\overline{p}_k
\leq 
(K_m K_{m'})^{r}\sum_{k=1}^{L-1}\binom{L-1}{k}\frac{1}{S^k}=(K_m K_{m'})^{r}\left[\left(1+\frac{1}{S}\right)^{L-1} -1\right].
\end{equation*}
\end{proof}

\begin{proof}[Proof of Lemma \ref{lemma:noisy_exchange_bound_TPE}]
The linear path $t\in[0,1]$ is discretised as $0=t^{}_1<\ldots<t^{}_L=1$, yielding the parameter vectors 
\begin{linenomath*}
\begin{equation*}
\param'(t):=
\begin{dcases}
(1-t)\dot{\param}+t\param'  & \text{if } d_m<d_{m'}, \\
(1-t)\param+t\param'        & \text{if } d_m=d_{m'}, \\
\end{dcases}
\end{equation*}
\end{linenomath*}
where $\dot{\param}$ is defined in Section \ref{section:ration_est}. The TPE requires $L-$1 sets of simulated data,
\begin{equation*}
y'=
\begin{matrix}
y^{(2)'}  & = &y^{(2)'}_1,&\ldots, &y^{(2)'}_S \sim f_{m'}(\cdot\mid \param'(t^{}_{2}))\\
\vdots&   &\vdots    &\ddots  &\vdots\\
y^{(L)'}  & = &y^{(L)'}_1,&\ldots, &y^{(L)'}_S \sim f_{m'}(\cdot\mid \param'(t^{}_{L})),
\end{matrix}
\end{equation*}
that are used to approximate the ratio of normalising constants using \eqref{eqn:PSE_ratio}. The remainder of this proof proceeds as in Lemma \ref{lemma:noisy_exchange_bound_SIS}. We check that
\begin{align*}
\mathbb{E}_{y'}\left|\hat{\rho}(x,x',y')-\rho(x,x')\right|
&\leq 
\int f(y'\mid x')\left|\hat{\rho}(x,x',y')-\rho(x,x')\right|{\rm d}y'\\
&=
\frac{q_{m'}(y\mid\param')}{q_m(y\mid\param)}
\frac{p_{m'}(\param')}{p_{m}(\param)}
\frac{p(\models_{m'})}{p(\models_m)}
\frac{h(\param,m\mid\param',m')}{h(\param',m'\mid\param,m)} \\
&\mathrel{\phantom{=}} \times
\mathbb{E}_{y'}\left|\prod_{j=1}^{L-1}\frac{1}{S}\sum_{b=1}^{S}\frac{q_m(y^{(j+1)'}_b\mid \param'(t^{}_j)))}{q_{m'}(y^{(j+1)'}_b\mid \param'(t^{}_{j+1}))} -\frac{z_{m}(\param)}{z_{m'}(\param')}\right| \\
&\leq 
\frac{q_{m'}(y\mid\param')}{q_m(y\mid\param)}
\frac{p_{m'}(\param')}{p_{m}(\param)}
\frac{p(\models_{m'})}{p(\models_m)}
\frac{h(\param,m\mid\param',m')}{h(\param',m'\mid\param,m)} \\
& \mathrel{\phantom{=}}
\times
\sqrt{
\mathbb{V}_{y'}
\left[
\prod_{j=1}^{L-1}\frac{1}{S}\sum_{b=1}^{S}\frac{q_m(y^{(j+1)'}_b\mid \param'(t^{}_j)))}{q_{m'}(y^{(j+1)'}_b\mid \param'(t^{}_{j+1}))}
\right]
} \\
&=
\Phi(x,x')\sqrt{\mathbb{V}_{y'}\left[\prod_{j=1}^{L-1}\frac{1}{S}\sum_{b=1}^{S}\frac{q_m(y^{(j+1)'}_b\mid \param'(t^{}_j)))}{q_{m'}(y^{(j+1)'}_b\mid \param'(t^{}_{j+1}))}\right]} \\
&=
\gamma^{}_{TP}(x,x').
\end{align*}
Applying Lemma \ref{lemma:var_prod_indep_rvs} leads to
\begin{align}\label{eq:gamma_TP_bound}
\gamma^{}_{TP}(x,x')
&\leq 
\Phi(x,x')[K_m K_{m'}]^{\frac{L-1}{2}}\left\{\left(1+\frac{1}{S}\right)^{L-1} -1\right\}^{1/2} \nonumber\\
&\leq
c^{}_{p_m}
c^{}_{p_{m'}}
c^{}_{\models_m}
c^{}_{\models_{m'}}
c^{}_{\upsilon(x,x')}
c^{}_{\upsilon(x',x)}
K_m K_{m'}
[K_m K_{m'}]^{\frac{L-1}{2}}\left\{\left(1+\frac{1}{S}\right)^{L-1} -1\right\}^{1/2}\nonumber\\
&=
c^{}_{p_m}
c^{}_{p_{m'}}
c^{}_{\models_m}
c^{}_{\models_{m'}}
c^{}_{\upsilon(x,x')}
c^{}_{\upsilon(x',x)}
[K_m K_{m'}]^{\frac{L+1}{2}}\left\{\left(1+\frac{1}{S}\right)^{L-1} -1\right\}^{1/2}.
\end{align}
\end{proof}

\begin{proof}[Proof of Theorem \ref{theorem:noisy_exchange_bound_TPE}]
Following the proof of Theorem \ref{theorem:noisy_exchange_bound_SIS}, it is straightforward to show that condition {\bf (H1)} in Corollary \ref{cor:noisy_rjmcmc_bound} is satisfied, with
\begin{equation*}
Q = 2 \text{ and }
\xi = 1 - \bigg[c^{}_{p_m}c^{}_{p_{m'}}
c^{}_{\models_m}
c^{}_{\models_{m'}}
c^2_{\upsilon(x,x')}
c^{}_{\upsilon(x',x)}
(K_m K_{m'})^{\frac{L+1}{2}}
\bigg]^{-1}.
\end{equation*}
Moreover, \eqref{eq:gamma_TP_bound} satisfies condition {\bf (H2)} in Corollary \ref{cor:noisy_rjmcmc_bound}. Applying this Corollary gives
\begin{equation*}
\sup_{x_0\in\mathcal{X}}\|\delta_{x_0}\mathcal{P}^n - \delta_{x_0}\hat{\mathcal{P}}^n\|
\leq 
\mathcal{D}^{}_{TP}\left\{\left(1+\frac{1}{S}\right)^{L-1} -1\right\}^{1/2},
\end{equation*}
with
\begin{equation}\label{eq:UB_TP}
\mathcal{D}^{}_{TP}=
c^{}_{p_m}
c^{}_{p_{m'}}
c^{}_{\models_m}
c^{}_{\models_{m'}}
c^{}_{\upsilon(x,x')}
c^{}_{\upsilon(x',x)}
[K_m K_{m'}]^{\frac{L+1}{2}}\left(\phi + \frac{Q\xi^{\phi}}{1-\xi}\right)
\end{equation}
and $\phi=\left\lceil \frac{\log(1/Q)}{\log(\xi)} \right\rceil$.
\end{proof}

\end{document}